\documentclass[%
 aip,jcp,superscriptaddress,
 amsmath,amssymb,floatfix,longbibliography,
 reprint,%
]{revtex4-2}
\usepackage{nicematrix}
\usepackage[utf8]{inputenc}
 \usepackage[T1]{fontenc}
\usepackage{chngcntr}

\usepackage{graphicx}
\usepackage{dcolumn}
\usepackage{bm}
\usepackage{xcolor}
\usepackage[normalem]{ulem}
\usepackage{physics}
\usepackage{chemformula}
\usepackage{soul}
\usepackage{subcaption}

\let\ket=\pket 
\let\bra=\pbra 
\newcommand{\ts}{\textsuperscript}

\usepackage{braket}
\usepackage{tikz}
\usetikzlibrary{external}
\usepackage{float}
\usepackage[colorlinks]{hyperref}
\hypersetup{ 
    colorlinks=true,
    linkcolor=red!80!black,
    urlcolor=magenta!80!black,
    citecolor=green!70!black,
}

\DeclareMathOperator*{\argmin}{arg\,min}
\begin{document}
\preprint{}

\title{Time evolution as an optimization problem: The hydrogen atom in strong laser fields in a basis of time-dependent Gaussian wave packets}

\author{Simon Elias Schrader}
\email{s.e.schrader@kjemi.uio.no}
\author{Håkon Emil Kristiansen}%
\author{Thomas Bondo Pedersen}%
\author{Simen Kvaal}%

\affiliation{Hylleraas Centre for Quantum Molecular Sciences, Department of Chemistry, University of Oslo, P.O. Box 1033 Blindern, N-0315 Oslo, Norway 
}%

\date{\today}

\begin{abstract}\small
    Recent advances in attosecond science have made it increasingly important to develop stable, reliable and accurate algorithms and methods to model the time evolution of atoms and molecules in intense laser fields. 
A key process in attosecond science is high-harmonic generation, which is challenging to model with fixed Gaussian basis sets, as it produces high-energy electrons, with a resulting rapidly varying and highly oscillatory wave function that extends over dozens of ångstr{\"o}m. Recently, Rothe's method, where time evolution is rephrased as an optimization problem, has been applied to the one-dimensional Schr{\"o}dinger equation. Here, we apply Rothe's method to the hydrogen wave function and demonstrate that complex-valued Gaussian wave packets with time-dependent width, center, and momentum parameters are able to reproduce spectra obtained from essentially exact grid calculations for high-harmonic generation with only 50-181 Gaussians for field strengths up to $5\times 10^{14}$W/cm$^2$. This paves the way for the inclusion of continuum contributions into real-time, time-dependent electronic-structure theory with Gaussian basis sets for strong fields, and eventually accurate simulations of the time evolution of molecules without the Born-Oppenheimer approximation. 
\end{abstract}

\maketitle

\section{Introduction}\label{sec:Introduction}
Founded on the discovery~\cite{HHG_discovery_1,HHG_discovery_2} and physical understanding~\cite{three_step_model,Lewenstein_model} of high-harmonic generation (HHG) in gases, the field of attosecond science has expanded rapidly in recent decades.~\cite{Corkum2007,Krausz2009,Lewenstein_QO_QI}
Attosecond laser pulses generated by HHG have paved the way for exciting experimental advances such as real-time observation of electron dynamics,~\cite{Fohlisch2005} attosecond imaging,~\cite{Haessler2010} and control of ionization dynamics in noble gases,~\cite{Hutten2018} and may eventually allow us to manipulate and control chemical reactions at the length- and time-scales of the electron.~\cite{Nisoli2017}
In 2023, Agostini, Krausz, and L'Huillier were awarded the Nobel Prize in Physics~\cite{NobelPrize} for their pioneering work that has made these developments possible.

In order to understand, interpret, and predict molecular processes initiated by attosecond laser pulses, one needs to solve the time-dependent Schr{\"o}dinger equation (TDSE).
This is made particularly challenging by the broad intensity distribution in the frequency domain of an attosecond laser pulse, which causes significant population of multiple electronic states, including the electronic continuum, and thus breaks the assumptions underpinning the adiabatic Born-Oppenheimer approximation.~\cite{Born1927,Born1954} 
Ideally, computational methods should be developed that treat electrons and nuclei on the same quantum-mechanical footing without invoking the Born-Oppenheimer approximation at any stage.
While such methods have been developed for solving the time-independent Schr{\"o}dinger equation for bound stationary states with very high accuracy,~\cite{Mitroy_Gaussian2013,Bubin_Adamowicz_ECG2013} virtually all time-dependent methods are based on the Born-Huang expansion~\cite{Born1954} of the molecular wave function in one form or other---see, e.g., Ref.~\citenum{Palacios2015}.

Methods to accurately solve the TDSE for the interaction between light and atoms using the single-active electron approximation (SAE)\cite{SAE_1,SAE_3} include spatial grid methods,~\cite{Bandrauk} the use of B-splines,~\cite{Bspline_2} time-dependent configuration interaction (TDCI) methods using large, tailored Gaussian basis sets,~\cite{Luppi,Coccia,Wozniak} and hybrid B-spline/Gaussian basis sets.~\cite{Bspline_1} Time-dependent methods to solve the TDSE for larger electronic systems include time-dependent density functional theory (TDDFT)~\cite{TDDFT} and wave function methods~\cite{Time_dependent_methods,Time_dependent_methods_2,Time_dependent_methods_3,HHG_methods_review} such as multiconfigurational time-dependent Hartree-Fock (MCTDHF) theory,~\cite{MCTDHF1,MCTDHF2} time-dependent coupled-cluster (TDCC) theory,~\cite{TDCC_review} TDCI methods,~\cite{TDCI} grid and B-spline methods,~\cite{Bspline_3} and hybrid methods.~\cite{Gauss_DVR,Gauss_Bspline} 
In the context of attosecond science, the key feature of these methods is that they allow for a description of ionization processes if the single-particle basis functions are suitably chosen.

Unfortunately, the atom-centered, isotropic, and usually real-valued Gaussian-type orbitals typically used in molecular electronic-structure theory~\cite{MEST} are not well suited as basis functions for the electronic continuum, implying that highly nonlinear optical phenomena such as ionization and HHG cannot be efficiently described.
Even if conventional Gaussian basis sets are augmented with Gaussians fitted to represent continuum (and Rydberg) functions,~\cite{Kaufmann} very large sets are needed to obtain reasonably converged HHG spectra.~\cite{Luppi,Coccia,Wozniak}
A rather obvious idea to capture the electronic continuum would be to multiply a conventional Gaussian with a plane wave, creating complex-valued basis functions reminiscent of London
orbitals,~\cite{London1937} which are routinely used for electronic-structure calculations in finite magnetic fields~\cite{Lange2012} thanks to efficient integral-evaluation algorithms.~\cite{tellgren_nonperturbative_2008,Reine2012,Irons2017}

With complex widths and freely moving momenta and centers, such basis functions become Gaussian wave packets, which form an overcomplete basis of the Hilbert space of square-integrable functions.
Gaussian wave packets are used as basis functions in, e.g., the variational multiconfigurational Gaussian (vMCG)~\cite{G-MCTDH,vMCG} approach to vibrational dynamics where the
nonlinear Gaussian parameters are determined using the time-dependent variation principle.
However, numerical instabilities are frequently encountered due to severe ill-conditioning of the Gramian matrix, which needs to be inverted to solve the variational equations of motion for the nonlinear Gaussian parameters,~\cite{Sawada,G-MCTDH,vMCG,KAY1989165,Lee2018,HaakonPhD} leading to unreliable results.
While there exist approaches to overcome these issues, they rely on assumptions about the potential and particle localization such as the local harmonic approximation,~\cite{Heller_1975} or they introduce approximations such as frozen/thawed Gaussians~\cite{Heller_1981,Vanicek2020} or independent propagation of Gaussians.~\cite{dutra_quantum_2020} 

Another drawback is that there is no natural way to enlarge the variational space, as there is no obvious way to remove or add Gaussian functions {(see however Ref. ~\citenum{Martinazzo_Burghardt} for a suggestion how this might be done)}.
This is particularly problematic for the simulation of ionization and HHG processes where the initial state is typically the electronic ground state, which is relatively localized with definite angular momentum, whereas the final state is highly delocalized and spread over numerous angular-momentum states.
An accurate yet efficient description of the full path from the initial to the final state thus requires an adaptive basis.
For this reason, grid methods are often preferred and can yield excellent results as long as the simulation box is large enough to avoid reflections from the boundaries. The reflections---which are also present in fixed-basis approaches---can be counteracted by a masking function,~\cite{Masking_functions} heuristic lifetime models,~\cite{Klinkusch,Coccia,Coccia2} or complex absorbing potentials {(CAPs)}.~\cite{Manolopoulos,Giovannini,Yu_2018}
Steep computational scaling with the number of degrees of freedom, especially when electron correlation is taken into account, makes grid methods intractable for larger systems, however.
A fully adaptive Gaussian basis does not require any such reflection countermeasures, although in practice one might employ them to keep computational effort at a manageable level. 
Compared with grid methods, a great advantage of the Gaussian-based approach is a very compact representation of the full wave-function history, and one disadvantage is the well-known inability of (a finite number of) Gaussians to reproduce the cusp conditions at electron-electron and electron-nucleus coalescence points.~\cite{MEST}

In this work, aiming at an accurate calculation of HHG spectra, we expand the wave function of a hydrogenic electron exposed to a strong $800\,\text{nm}$ laser pulse in a basis of complex-valued Gaussian wave packets.
As demonstrated in a recent proof-of-principle study, relatively few Gaussian wave packets are required to capture both electronic and rovibrational dynamics induced by strong laser pulses provided that the number of Gaussians as well as their nonlinear parameters are adjusted in each time step.~\cite{Wozniak_Gaussians}
Hence, to achieve full spatio-temporal adaptivity, we treat both the number of Gaussians and \emph{all} linear and nonlinear parameters as time-dependent variables.
Avoiding the numerical instabilities alluded to above, we use Rothe's method~\cite{Rothe1930_MA} to recast the evolution equation as an optimization problem at each time step.
Rothe's method has previously been applied to study a similar one-dimensional hydrogenic problem,~\cite{kvaal2023need}
to propagate neural-network quantum states,~\cite{Gutierrez2022} and to solve the quantum-classical Liouville equation.~\cite{Horenko2004adaptive}

The paper is organized as follows. In Sec.~\ref{sec:Methods}, we first describe the details of Rothe's method and how it rephrases time evolution as an optimization problem.
This is followed by a description of the Gaussian basis set, and how the error in Rothe's method is related to the variance of the Hamiltonian. 
We then describe how the optimization is carried out, with particular focus on adding and removing Gaussian basis functions.
In Sec.~\ref{sec:Results}, we present HHG spectra produced using Rothe's method and compare them to HHG spectra obtained using the discrete variable representation and Kaufmann basis functions~\cite{Kaufmann} with a heuristic lifetime model.
We conclude in Sec.~\ref{sec:Conclusion} with a discussion of the strengths and shortcomings of Rothe's method, and how it might be used for modelling larger systems.

\section{Methods}\label{sec:Methods}
\subsection{Rothe's method for the time-dependent Schr{\"o}dinger equation}\label{sec:Rothe}
Using atomic units throughout and suppressing the dependence of the wave function on spatial coordinates for notational convenience, the TDSE may be written as
\begin{equation}
    \text{i}\frac{\partial}{\partial t}\Psi(t)=\hat{H}(t)\Psi(t).
\end{equation}
It is possible to reformulate the TDSE variationally as 
\begin{equation}\label{eq:VarTid}
    \frac{\partial}{\partial t}\Psi(t)=\text{arg}\min_\omega\norm{\hat{H}(t)\Psi(t)-\text{i}\omega}^2.
\end{equation}
This formulation leads to the McLachlan variation principle,~\cite{McLachlan} which is equivalent to the more widely used Dirac-Frenkel variation principle~\cite{Dirac1930,Frenkel_wave} under certain conditions.~\cite{broeckhove_equivalence_1988} {When invoking the Dirac-Frenkel or McLachlan variation principle for approximate dynamics, the time-dependent wave function is restricted to lie in a time-independent manifold $\mathcal{M}\subset L^2$, which depends on the parameterization of the wave function ansatz.}
In the following, a different approach is used, where time evolution is reformulated as an optimization problem. Our derivation here follows closely the one given by \citeauthor{kvaal2023need}\cite{kvaal2023need} and differs only in details regarding the discretization scheme used.
The Crank-Nicolson scheme for the TDSE reads\cite{joachain}
\begin{equation}
\label{eq:crank-nicolson}
    \hat A_{i}\Psi(t_{i+1})= \hat A_{i}^\dagger\Psi\left(t_i\right),
\end{equation}
where
\begin{equation}
    \hat A_{i}=\hat I+\text{i}\frac{\Delta t}{2}\hat H\left(t_i+\frac{\Delta t}{2}\right),
\end{equation}
and $t_{i+1}=t_{i}+\Delta t$ with $\Delta t$ denoting the time step. Here, $\hat I$ is the identity operator.
A variational formulation then reads 
\begin{equation}\label{eq:Rothe}
    \Psi(t_{i+1})=\text{arg}\min_{\chi}\norm{\hat A_i\chi-\hat A_i^\dagger\Psi(t_i)}^2.
\end{equation}
While resembling eq. \eqref{eq:VarTid}, eq. \eqref{eq:Rothe} explicitly solves for the \textit{wave function} at the next time step, not the \textit{time derivative}. This is Rothe's method, or the method of horizontal lines, also referred to as method of adaptive time layers. \cite{deuflhard2012adaptive} 
While one cannot hope to solve eq. \eqref{eq:Rothe} exactly, it can be solved approximately by choosing a particular representation for $\chi(t)$. 
Any representation that makes it possible to approximately solve eq. \eqref{eq:Rothe} can be chosen such as, e.g., a real-space discretization or an expansion in a basis. We discuss the case where $\chi(t)$ is written as a linear combination of parameter-dependent functions at every time step
\begin{equation}\label{eq:WF_ansatz}
    \chi(t)=\sum_{m=1}^{M(t)} c_m(t)\phi_m(\boldsymbol{\alpha}(t)),
\end{equation}
where the basis functions $\{\phi_1(\boldsymbol {\alpha} (t)),\dots,\phi_{M(t)}(\boldsymbol \alpha(t))\}$ may be nonlinear in the parameters $\boldsymbol{\alpha}(t)$. Inserting \eqref{eq:WF_ansatz} for $\chi$ in eq. \eqref{eq:Rothe}, the problem of finding the best wave function corresponds to finding the optimal parameters $\boldsymbol{\alpha}^{opt.}_{i+1},\boldsymbol{c}^{opt.}_{i+1}$ describing $\Psi(t_{i+1})$ 
\begin{equation}\label{eq:Rothe_opt}
\boldsymbol{\alpha}^{opt.}_{i+1},\boldsymbol{c}^{opt.}_{i+1}=\text{arg}\min_{\boldsymbol{\alpha},\boldsymbol{c}}\norm{\sum_{m=1}^{M(t)}c_m\check \phi_m(\boldsymbol{\alpha})-\Tilde{\Psi}_i}^2,
\end{equation}
where we have introduced the notation
\begin{align}
&\Tilde{\Psi}_i = \hat A_i^\dagger\Psi(t_i), \\
&\check \phi_m(\boldsymbol{\alpha}) = \hat A_i \phi_m(\boldsymbol{\alpha}).
\end{align}
To reduce the number of free parameters and to counteract ill-conditioning, Golub and Pereyra's Variable Projection (VarPro) algorithm \cite{Golub_Pereyra} can be employed to remove the linear parameters. Defining the matrix $\boldsymbol{S}^{i+1}(\alpha)$ and the vector $\boldsymbol{\rho}^{i+1}(\alpha)$ with elements
\begin{align}
    &S^{i+1}_{mn}(\boldsymbol{\alpha}) =\braket{\check \phi_m(\boldsymbol{\alpha})|\check \phi_n(\boldsymbol{\alpha})},\\
    &\rho^{i+1}_{m}(\boldsymbol{\alpha}) =\braket{\check \phi_m(\boldsymbol{\alpha})|\tilde\Psi_i},
\end{align}
the linear coefficients that optimize $\boldsymbol{c}$ for a given set of nonlinear parameters $\boldsymbol{\alpha}$ in eq. \eqref{eq:Rothe_opt} are given by
\begin{equation}
    \boldsymbol{c}_{i+1}(\boldsymbol{\alpha})= \boldsymbol{S}^{i+1}(\boldsymbol\alpha)^{-1}\boldsymbol{\rho}^{i+1}(\boldsymbol \alpha).
\end{equation}
Inserting this expression into eq. \eqref{eq:Rothe_opt}, the optimization is now with respect to $\boldsymbol{\alpha}$ only, reading
\begin{equation}
\boldsymbol{\alpha}^{opt.}_{i+1}=\text{arg}\min_{\boldsymbol{\alpha}}\left[\text{r}_{i+1}(\boldsymbol{\alpha})\right],
\end{equation}
where the \textit{Rothe error} is given by
\begin{align}  
    \text{r}_{i+1}(\boldsymbol{\alpha})&=\norm{\sum_{m=1}^{M(t)}c_m(\boldsymbol{\alpha})\check \phi_m(\boldsymbol{\alpha})-\Tilde{\Psi}_i}^2 \nonumber \\
    &=\braket{\tilde\Psi_i|\tilde\Psi_i} \nonumber \\
    &- \boldsymbol\rho^{i+1}(\boldsymbol{\alpha})^\dagger \boldsymbol{S}^{i+1}(\boldsymbol\alpha)^{-1}\boldsymbol\rho^{i+1}(\boldsymbol{\alpha}).   
 \end{align}
The last term is readily recognized as the inner product of $\ket{\tilde\Psi_i}$ with itself projected onto the space spanned by the vectors $\check \phi_i(\boldsymbol\alpha)$.~\cite{Projector_theory} That is, defining the projector
\begin{equation}
    \hat{P}_{i+1}(\boldsymbol \alpha)=\sum_{mn}\ket{\check \phi_m(\boldsymbol \alpha)}[\boldsymbol{S}^{i+1}(\boldsymbol{\alpha})]^{-1}_{mn}\bra{\check \phi_n(\boldsymbol \alpha)},
\end{equation}
we find
\begin{align}
   \boldsymbol\rho^{i+1}(\boldsymbol{\alpha})^\dagger\boldsymbol{S}^{i+1}(\boldsymbol\alpha)^{-1}\boldsymbol\rho^{i+1}(\boldsymbol{\alpha})
   = \bra{\tilde\Psi_i}\hat{P}_{i+1}(\boldsymbol \alpha)\ket{\tilde\Psi_i}.
\end{align}
Observe that the calculation of the Rothe error requires the calculation of matrix elements of the form $\braket{\phi_m(\boldsymbol\alpha)|\phi_n(\boldsymbol\alpha)}$, $\braket{\phi_m(\boldsymbol\alpha)|\hat{H}(t)|\phi_n(\boldsymbol\alpha)}$ and $\braket{\phi_m(\boldsymbol\alpha)|\hat{H}^2(t)|\phi_n(\boldsymbol\alpha)}$. Analytical expressions for the overlap matrix, the Hamiltonian matrix and the squared Hamiltonian matrix can in many cases be derived with Gaussian basis functions.~\cite{Boys1960,Singer1960} 

As Rothe's method requires the calculation of the expectation values involving the squared Hamiltonian $\hat{H}^2(t)$, it is important that the basis functions $\ket{\phi_m(\boldsymbol\alpha)}$ all lie in the {\textit{domain} of $\hat{H}(t)$}, which is identical to the \emph{form domain} of $\hat{H}^2(t)$.~\cite{Reed_Simon} This requirement is equivalent to $\ket{\hat{H}(t)\phi_m(\boldsymbol\alpha(t))} \in L^2$ for all $m$ and all $t$. This is a stricter requirement than needed for the time-independent variation principle, which only requires that the basis functions lie in the form domain of $\hat{H}(t)$.

\subsection{Error control}
There are two sources of error in Rothe's method. The first is the Crank-Nicolson scheme, which has a global error that scales as $O((\Delta t)^2)$. The second is the fact that the Crank-Nicolson scheme is approximated, as the Rothe error will not be zero at every time step. Full error control is nevertheless possible. Defining a global error parameter $\varepsilon_{\text{max}}$, the maximal acceptable {square root of the} Rothe error at each time step is chosen as $\varepsilon_{\Delta t}=\varepsilon_{\text{max}}\Delta t / T_{f}$ where $T_f$ is the final time of the simulation. If, at a given time step, it is not possible to get the Rothe error below $\varepsilon_{\Delta t}^2$ by optimizing the existing parameters, that is, if $r_{i+1}(\mathbf{\alpha})>\varepsilon_{\Delta t}^2$, additional basis functions can be added to reduce the Rothe error below the chosen tolerance.

The Rothe error overestimates the actual deviation from the Crank-Nicolson scheme. Going from time step $t_i$ to $t_{i+1}$, the error $\varepsilon_{i+1}$ in Rothe's method compared to the exact Crank-Nicolson wave function is upper bounded by the Rothe error:
\begin{align}
        \varepsilon_{i+1}^2&=\left\|\Psi(t_{i+1})-(\hat A_{i})^{-1}\hat A_{i}^\dagger\Psi\left(t_i\right)\right\|^2 \nonumber \\
        &= \left\|\hat A_{i}^{-1}\left(\hat A_{i}\Psi(t_{i+1})-\hat A_{i}^\dagger\Psi\left(t_i\right)\right)\right\|^2 \nonumber \\
        &\leq \left\|\hat A_{i}^{-1}\right\|^2\left\|\hat A_{i}\Psi(t_{i+1})-\hat A_{i}^\dagger\Psi\left(t_i\right)\right\|^2 \nonumber \\
        &= \left\|\hat A_{i}^{-1}\right\|^2{r_{i+1}} \leq {r_{i+1}},
\end{align}
where $\Psi(t_{i+1})$ is the Rothe approximation to $(\hat A_{i})^{-1}\hat A_{i}^\dagger\Psi\left(t_i\right)$, and we used the fact that $\|A_i\|\geq 1$ which follows from the hermiticity of the Hamiltonian.

If the Crank-Nicolson scheme were exact, the total error in the wave function would then be bounded by
\begin{equation}
    \norm{\Tilde\Psi(T_f)-\Psi{(T_f)}}\leq \varepsilon_{\text{tot}} = \sum_{i=1}^{N_T} \sqrt{r_i},
\end{equation}
where $N_T$ is the number of time steps, $\Tilde\Psi(T_f)$ is the exact Crank-Nicolson wave function, and $\Psi{(T_f)}$ is the Rothe approximation. By reducing $\Delta t$ and $\varepsilon_{\text{max}}$, full control of the time evolution error is possible in principle. It should be mentioned, however, that the Rothe error may significantly overestimate the true time evolution error. In particular, it is possible that $\|\hat{A}_i\Psi(t)\|\gg \|\Psi(t)\|$. The reason for this is that $\hat{A}_i$ contains the Hamiltonian, which might weigh particular regions of the wave function more strongly. As an illustration, consider a particle interacting with a linearly polarized laser pulse in the dipole approximation in length gauge. The Hamiltonian is given by
\begin{equation}
    \hat H(t)=\hat H_0 + E(t)z,
\end{equation}
where $\hat H_0$ is the field-free Hamiltonian and $E(t)$ is the electric-field strength of the laser.
The term $\|z\Psi(t)\|$ becomes large as the wave function spreads far away from the origin, which is bound to occur for field strengths strong enough to induce ionization dynamics. From this point of view, the Rothe error {might significantly overestimate the time evolution error. Nevertheless, by reducing the Rothe error, we reduce an upper bound for the time evolution error, which leads to a probable reduction of the time evolution error itself.} In a similar fashion, the time-independent variation principle is a standard method used in quantum chemistry and physics to find approximation of the ground state of a Hamiltonian $\hat{H}$. Letting $\ket{\Phi}$ be the variational solution and $\ket{\Psi}$ the real ground state, $|{\braket{\Psi|\Phi}}|$ can be large even though $\bra{\Phi}\hat{H}\ket{\Phi}\gg\bra{\Psi}\hat{H}\ket{\Psi}$ (or vice versa - one can have $\bra{\Phi}\hat{H}\ket{\Phi}\approx\bra{\Psi}\hat{H}\ket{\Psi}$ while $|{\braket{\Psi|\Phi}}|$ is small). Reducing the variational energy, hence, does not guarantee that one gets close to the correct ground state, but it is nevertheless a reliable method in practice to obtain good approximations.

\subsection{Choice of basis set}
In principle, any basis set can be chosen for Rothe's method for which $\braket{\phi_m(\boldsymbol\alpha)|\phi_n(\boldsymbol\alpha)}$, $\braket{\phi_m(\boldsymbol\alpha)|\hat{H}(t)|\phi_n(\boldsymbol\alpha)}$ and $\braket{\phi_m(\boldsymbol\alpha)|\hat{H}^2(t)|\phi_n(\boldsymbol\alpha)}$ and their derivatives with respect to $\boldsymbol{\alpha}$ can be calculated effectively. Here, we opt for normalized, shifted-center Gaussians with complex width and shift as our basis because of their variability, their completeness properties, and the fact that most matrix elements of interest and their derivatives can be calculated analytically, which includes most of those involving $\hat{H}^2(t)$. That is, each basis element is parameterized as
\begin{equation}    \phi_m(\boldsymbol{\alpha})=\frac{g_m(\boldsymbol{\alpha})}{\sqrt{\left\langle g_m(\boldsymbol{\alpha})|g_m(\boldsymbol{\alpha})\right\rangle}},
\end{equation}
where $g_m(\boldsymbol{\alpha})$ is a non-normalized Gaussian with shifted center
\begin{equation}
g_m(\boldsymbol{\alpha})=\exp\left(-\left(\boldsymbol{r}-\boldsymbol{\mu}_m\right)^T\boldsymbol{A}_m\left(\boldsymbol{r}-\boldsymbol{\mu}_m\right)\right),
\end{equation}
where $\boldsymbol{A}_m$ is a complex-symmetric $n\times n$ matrix with $\Re(\boldsymbol{A}_m)>0$, which guarantees that $g_m(\boldsymbol{\alpha})$ is square-integrable, and $\boldsymbol \mu_m$ is a complex vector of size $n$, where $n$ is the dimensionality of the system considered.  Depending on the form of the Hamiltonian, it might be advantageous to introduce restrictions on $\boldsymbol{A}_m$ or $\boldsymbol{\mu}_m$. For example, for Coulomb potentials, analytic expressions for electron repulsion integrals and nuclear attraction integrals exist \cite{cafiero2001analytical} when
\begin{equation}
    \boldsymbol{A}_m=\boldsymbol{A}_m'\otimes I_3,
\end{equation}
where $\otimes$ denotes the Kronecker product.

As the parameter vector $\boldsymbol{\alpha}$ is time-dependent, our ansatz is very different from the case where only the linear coefficients depend on time, i.e. an ansatz of the form
\begin{equation}
    \Psi(t)=\sum_{m=1}^{M(t)} c_m(t)\phi_m.
\end{equation}
with time-independent Gaussians $\phi_m$.

The basis functions $\phi_m(\boldsymbol{\alpha})$ can have an arbitrary center and are not fixed at, e.g., an atomic nucleus, and they should \emph{not} be understood as (products of) Gaussian-type orbitals but rather as explicitly correlated, time-dependent wave packets.

\subsection{Time evolution and variance}
The \textit{variance} of the Hamiltonian for a given wave function $\ket{\Psi}$ is defined by
\begin{equation}
\text{Var}(\hat{H};\Psi)=\braket{\Psi|\hat{H}^2|\Psi}-\braket{\Psi|\hat{H}|\Psi}^2.
\end{equation}
The variance vanishes for an eigenstate of the Hamiltonian.
The Rothe error is related to the variance. Let $\ket{\Psi}$ be the variationally optimized approximate ground state of the time-independent Hamiltonian $\hat{H}$ in some orthogonal, finite-dimensional basis, with energy $E = \braket{\Psi|\hat{H}|\Psi}$.
Then $\ket{\Psi}$ will be the exact ground state of the projected Hamiltonian $\hat{H}_P=\hat{P}\hat{H}\hat{P}$, where $\hat{P}$ is an orthogonal projection on the space spanned by the basis. Considering the time evolution of $\ket{\Psi(0)}=\ket{\Psi}$, the time-dependent variation principle gives the solution $\ket{\Psi(t)}=e^{-\text{i}Et}\ket{\Psi}$. It is hence meaningful to consider the Rothe error in a basis consisting only of $\ket \Psi$. I.e., the Rothe error when insisting that $\ket{\Psi(t)}=c(t)\ket{\Psi}$ for some complex number $c(t)$. The corresponding projector therefore reads $\hat{P}=\ket{\Psi}\!\bra{\Psi}$. Letting $\ket{\tilde\Psi}=\hat{A}^\dagger\ket{\Psi}$, $\ket{\check\Psi}=\hat{A}\ket{\Psi}$, we find the $\boldsymbol{\alpha}$-independent and time-independent Rothe error
\begin{equation}
    r=\braket{\Tilde{\Psi}|\Tilde{\Psi}}-\frac{\abs{\braket{\tilde{\Psi}|\check{\Psi}}}^2}{\braket{\check \Psi|\check \Psi}}=(\Delta t)^2\text{Var}(\hat{H};\Psi)+O((\Delta t)^4),
\end{equation}
which can be obtained by Taylor expansion in $\Delta t$. Hence, the variance and Rothe error are directly related. The variance also shows up in error estimates for the Dirac-Frenkel variation principle.~\cite{Martinazzo_Burghardt, lubich2008quantum} \citeauthor{lubich2008quantum}~\cite{lubich2008quantum} gives the error due to time evolution as 
\begin{equation}
    \|\Psi(t)-\tilde\Psi(t)\|\leq \int_0^t \|\frac{d\Psi}{dt}(s)+\text{i}\hat{H}\Psi(s)\|\ \text{d}s,
\end{equation}
where $\ket{\tilde\Psi(t)}$ is the exact wave function stemming from the time evolution of $\ket{\Psi(0)}=\ket{\tilde\Psi(0)}$ and $\ket{\Psi(t)}$ the approximation to the time-evolved wave function. If the Hamiltonian is time-independent, the variational solution is $\ket{\Psi(t)}=\text{e}^{-\text{i}Et}\ket{\Psi(0)}$, and the error becomes
\begin{align}
    &\|\Psi(t)-\tilde\Psi(t)\|^2 \leq \nonumber \\
    &\left(\int_0^t \|-\text{i}E\text{e}^{-\text{i}Es}\Psi(0)+\text{i}\hat{H}\text{e}^{-\text{i}Es}\Psi(0)\|\ \text{d}s\right)^2
    \nonumber \\
    &=t^2{\text{Var}(\hat{H};\Psi(0))}.
\end{align}
The relation between time-evolution error and variance hence is not restricted to Rothe's method, but is inherent to time-dependent methods.

\subsection{Regularization of the Coulomb potential}
The Hamiltonian operator for the hydrogen atom reads
\begin{equation}\label{eq:exactH}
    \hat{H}=-\frac{1}{2}\nabla^2-\frac{1}{r}.
\end{equation}
It is a well-known fact that the exact eigenfunctions of Coulombic Hamiltonians need to fulfill Kato's cusp condition\cite{Kato_cusp} (and similar higher-order cusp conditions)~\cite{pack_cusp_1966,Kurokawa_cusp,Savin_cusp} which stems from the fact that the Coulomb singularity needs to be exactly ``cancelled'' by the kinetic-energy operator. The ground-state wave function of the hydrogen atom,
\begin{equation}
    \Psi_0(\boldsymbol{r}) = \frac{1}{\sqrt{\pi}} \text{e}^{-r},
\end{equation}
cannot be exactly written as a linear combination of a finite number of Gaussians, as $\Psi_0(\boldsymbol{r})$ is not smooth at $\boldsymbol{r} = \boldsymbol{0}$ while any finite linear combination of Gaussians will be smooth. The inability of Gaussians to exactly reproduce the correct behavior of the exact wave function at the origin applies to other atoms and molecules as well, where all nuclear-electron and electron-electron cusps need to be taken into consideration. It also applies to those single-active-electron (SAE) potentials that behave like $Z/r$ as $r\to 0$.~\cite{Schiessl}

The variance is highly sensitive to the presence of singular operators in the Hamiltonian, causing slow Gaussian basis-set convergence,~\cite{Ireland_lower_bounds} and leading to large Rothe errors that can only be reduced by adding vast numbers of narrow Gaussians. This is unfortunate if the property of interest---say, the HHG spectrum---does not depend strongly on the wave function near the origin but rather on its long-range tail. For this reason we now consider the regularized Hamiltonian for the hydrogen atom
\begin{equation}\label{eq:approxH}
    \hat{H} = -\frac{1}{2}\nabla^2-\frac{\erf(\mu r)}{r}, \quad\text{$\mu>0$}.
\end{equation}
The potential $\erf (\mu r)/r$ is nonsingular and, hence, no cusp conditions apply.
Observe that $\erf(\mu r)/r \to 1/r$ pointwise for $r\in(0,\infty)$ as $\mu\to \infty$. This type of regularization is used in the context of range separation.~\cite{range_separation1,range_separation2,range_separation3} It is also used in relativistic quantum chemistry calculations where the non-zero radius of the nucleus is taken into account, as it solves the Poisson equation for a Gaussian charge density.~\cite{visscher1997dirac} A possible, related alternative is the erfgau potential, defined for the hydrogen atom as
\begin{equation}
    V_e(r)=-\left(\frac{\erf(\mu r)}{r}+c_e\exp(-(\alpha_e r)^2)\right),
\end{equation}
where~\cite{range_separation1}
\begin{equation}
 c_e=0.923+1.568 \mu, \qquad
 \alpha_e=0.2411+1.405 \mu.
\end{equation}
The erfgau potential has a spectrum very similar to that of the exact Coulomb potential, and it has recently been successfully applied to the time evolution of molecular systems and the HHG process in particular by \citeauthor{Orimo2023}~\cite{Orimo2023} Although there are some theoretical arguments that challenge the applicability of softened potentials for the HHG process,~\cite{Role_of_Coulomb_HHG} the study by \citeauthor{Orimo2023} as well as the present work show that this is not a problem in practice for the field strengths considered. 

\begin{figure}[H]
    \centering
    \includegraphics[width=0.5\textwidth]{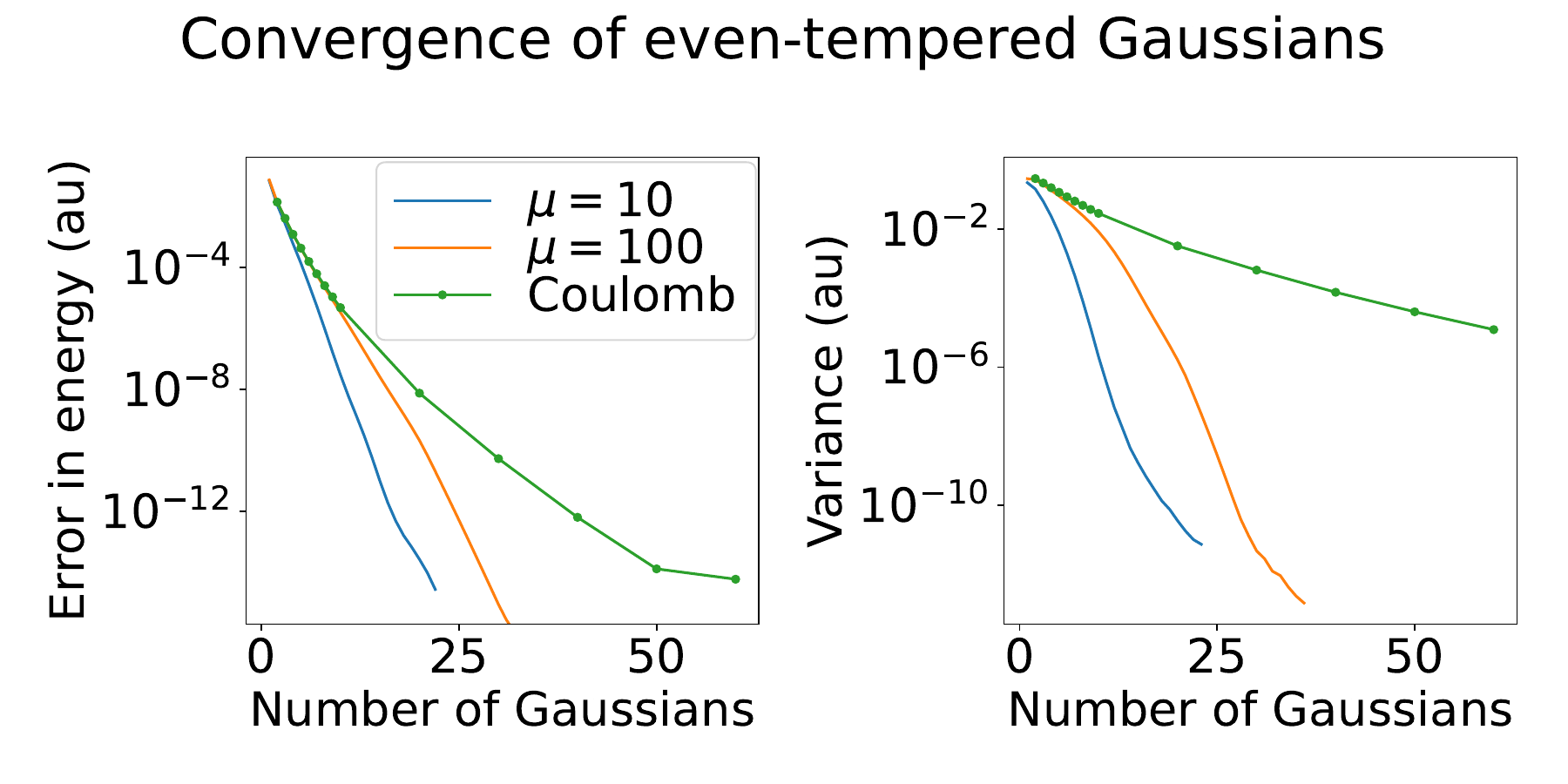}
    \caption{Error in energy (left) and variance (right) as functions of the number of Gaussians in an even-tempered basis set for a regularized Coulomb potential ($\mu\in\{10,100\}$) and for the exact Coulomb potential. As the exact energy of the ground state is not available for the regularized Coulomb potential, we used the energy found for $N=23$ Gaussians ($\mu=10$) and $N=37$ Gaussians ($\mu=100$) as a reference. Parameters for the exact Coulomb potential are taken from Ref. \citenum{Bakken2004}. Parameters for the regularized Coulomb potential were obtained by minimizing the energy.}
    \label{fig:Convergence_gauss}
\end{figure}

The fact that linear combinations of Gaussians do not show the correct behavior near the origin for the exact Hamiltonian leads to a large variance. Even with a linear combination of 60 Gaussians in an even-tempered basis set, which accurately reproduces the ground state energy of the hydrogen atom $E=-0.5\,\text{Ha}$ with an error below $10^{-15}\,\text{Ha}$,~\cite{Bakken2004} the variance of this variationally optimized wave function remains above $10^{-5}\,\text{Ha}^2$. The slow convergence of the variance has also been observed in a basis of explicitly correlated Gaussians for small molecules, even when the variance is optimized instead of the energy.~\cite{Ireland_lower_bounds} Similar challenges regarding the slow convergence of particular expectation values when using Gaussian basis sets, some of which are necessary to evaluate the squared Hamiltonian, also arise in the calculation of quantum electrodynamical contributions, such as the Bethe logarithm.~\cite{Adamowicz_Bethe} Regularizing the Coulomb potential leads to a much faster convergence of the variance. Figure \ref{fig:Convergence_gauss} shows the convergence of the energy and variance as function of the number of basis functions when using an even-tempered, atom-centered Gaussian basis set to approximate the ground state for the exact (eq. \eqref{eq:exactH}) and the regularized (eq. \eqref{eq:approxH}) Coulomb potential. It can be seen that regularization of the Coulomb potential significantly reduces the number of Gaussians required for energy and variance convergence, and it is possible to obtain a variance below $10^{-10}\,\text{Ha}$ with $19$ and $25$ Gaussians for $\mu = 10$ and $\mu = 100$, respectively. 

These arguments show that regularization of the Coulomb potential is necessary to have a small Rothe error. For time-dependent phenomena, without regularization, it might happen that the Rothe error at a given time step would be minimized below a given tolerance by focusing on the behavior near the cusp, thus missing the dynamics of interest.
While Rothe's method allows us to explicitly take into account finite-basis effects and inexact evolution, it is also extremely sensitive to effects which are due to the inherent inability of a reasonably sized Gaussian basis set to describe the cusp. It should be mentioned, however, that a feasible alternative to regularization would be to use integral-transformation techniques to accelerate convergence of expectation values as described in Refs.~\citenum{IT_1} and \citenum{IT_2}.

\subsection{The time-dependent Hamiltonian and choice of basis}
In this study, we consider a single hydrogen atom with a regularized Coulomb potential interacting with a laser pulse in the electric-dipole approximation. The Hamiltonian
is given by
\begin{equation}\label{eq:Hamiltonian}
    \hat{H}(t)=-\frac{1}{2}\nabla^2-\frac{\erf(\mu r)}{r}+E_0f(t)\sin(\omega t)z,
\end{equation}
where $\omega$ is the carrier frequency of the laser pulse, $E_0$ the peak electric-field strength {in the absence of an envelope function}, and $f(t)$ is an envelope function.
Following \citeauthor{Barth2009},~\cite{Barth2009} we use a trigonometric envelope which is nonzero only in a finite time interval, i.e.,
\begin{equation}
  f(t) = \sin^2\!\left(\frac{\pi t}{T_f}\right), \qquad 0 \leq t \leq T_f,
\end{equation}
where $T_f = 2\pi N_c / \omega$ and $N_c$ is the number of optical cycles.

The basis functions are complex-valued, isotropic Gaussian wave packets, i.e.,
\begin{align}
&g_m(\boldsymbol{r},t) = g(\boldsymbol{r}; \boldsymbol{\alpha}_m(t)) \nonumber \\
&= \exp\left(-(a_m^2+\text{i}b_m)\norm{\boldsymbol{r}-\boldsymbol{\mu}_m}^2+\text{i}\boldsymbol{q}_m^T(\boldsymbol{r}-\boldsymbol{\mu}_m)\right),
\end{align}
where $\boldsymbol \mu_m=(0,0,\mu_m)^T$ and $\boldsymbol q_m=(0,0,q_m)^T$ due to the cylindrical symmetry of the Hamiltonian. Each parameter $a_m(t),b_m(t),q_m(t),\mu_m(t)$ is real, and the parameter vector $\boldsymbol{\alpha}$ is defined as
\begin{align}
\boldsymbol{\alpha}(t) = \big(&a_1(t),b_1(t),\mu_1(1),q_1(t),\dots, \nonumber \\
                              &a_{M(t)}(t),b_{M(t)}(t),\mu_{M(t)}(t),q_{M(t)}(t)\big)^T.
\end{align}

\subsection{Optimization of nonlinear parameters}
As has been pointed out in the context of variational optimization of the nonlinear parameters of explicitly correlated Gaussians for ground and excited states,~\cite{Mitroy_Gaussian2013} it is important to maintain a reasonable balance between the effort spent on the optimization procedure and keeping the number of basis functions as low as possible. Intense optimization is time-consuming, but so is using a larger basis than necessary for a given error threshold $\varepsilon_{\Delta t}$.

To optimize the nonlinear parameters, we use the SciPy~\cite{scipy} implementation of the Broyden-Fletcher-Goldfarb-Shanno (BFGS) algorithm~\cite{nocedal_numerical} with analytical derivatives of the Rothe error with respect to all parameters.
As an initial guess for the inverse of the Hessian matrix, we used $10^8$ times the identity matrix.
We found that this gave improved results compared to just using the identity matrix, which is due to the fact that the initial guess for the inverse of the Hessian in the BFGS algorithm is not scale invariant.
We use SymPy~\cite{Sympy} to calculate the derivatives of the overlap matrix, the Hamiltonian matrix and the squared Hamiltonian matrix with respect to the nonlinear parameters.  As initial parameters for every time step, we use the parameters from the previous time step plus some of the numerical derivatives from the previous change, i.e.,  $\boldsymbol{\alpha}_{i+1}^{\text{init.}}=\boldsymbol{\alpha}_i+d(\boldsymbol{\alpha}_i-\boldsymbol{\alpha}_{i-1})$, where $d\in[0,1.2]$ is determined by a line search.
If the number of parameters has changed going from time step $i-1$ to time step $i$, we pad $\boldsymbol{\alpha}_{i-1}$ with zeros or remove the last elements to match its length with that of $\boldsymbol{\alpha}_{i}$.
{When the optimization algorithm converges in less than 50 iterations and the Rothe error at a time step is bigger than the threshold, i.e., $r_{i+1}(\mathbf{\alpha})>\varepsilon_{\Delta t}^2$, the optimization is re-run with slightly perturbed initial parameters.}
If the Rothe error is still higher than the threshold, we assume that no feasible minimum can be found using the amount of basis functions currently used, and up to two additional basis functions are added.

We have observed that optimization between time steps can be accelerated and stabilized by freezing the nonlinear parameters of the Gaussians representing the ground state, which we have done in all calculations.
Although this procedure may increase the total number of basis functions, it is justified for the simulation of HHG processes, as overlap of the wave function with the ground state is expected to remain large, especially in the beginning of the simulation and for the smallest field strengths considered.
Furthermore, we have observed that putting bounds on the range for all nonlinear parameters simplifies the optimization procedure and can give large speedups (without constraints, there are $M(t)!$ global minima for every time step, as relabeling basis functions has no effect on the wave function, and each local minimum is also $M(t)!$-fold degenerate). In particular, the $j$th nonlinear parameter $(\boldsymbol{\alpha}_{i+1})_{j}$ is only allowed to change by some number $p$ and a small extra step $q$: 
\begin{equation}
\begin{split}
    {\min}_j&=\boldsymbol{(\alpha}_{i+1}^{\text{init.}})_j-p|\boldsymbol{(\alpha}_{i+1}^{\text{init.}})_j|-q\\
    {\max}_j&=\boldsymbol{(\alpha}_{i+1}^{\text{init.}})_j+p|\boldsymbol{(\alpha}_{i+1}^{\text{init.}})_j|+q\\
    (\boldsymbol{\alpha}_{i+1})_{j}&\in[{\min}_j,{\max}_j].
\end{split}
\end{equation}
This is done by performing optimization with respect to a new set of parameters $\boldsymbol{\alpha}'_{i+1}$
\begin{equation}
     (\boldsymbol{\alpha}'_{i+1})_j  = {{\min}_j} + \big(\tanh( (\boldsymbol{\alpha}'_{i+1})_j)\! +\! 1\big) \frac{({\rm {\max}_j} - { {\min}_j})}{2}.
 \end{equation}
We have chosen $p=0.5$ and $q=0.1$ for all field strengths considered. These parameters were chosen to avoid that additional basis functions are added prematurely, though different choices of basis functions or different systems might require different choices of parameters $p,q$, or they might not require constraints altogether. 
To avoid near-linear dependence in the basis, we use the penalty scheme described in Ref.~\citenum{Bubin_Adamowicz_ECG2013}. 
Specifically, {the function that is minimized with respect to $\boldsymbol{\alpha}$ is the Rothe error plus an additional} penalty term $\sum_{mn} \mathcal{P}_{mn}$ where
\begin{equation}
\label{eq:penalty}
\mathcal{P}_{mn}=
\begin{cases}
    \beta \frac{\left|\braket{\phi_m|\phi_n}\right|^2-0.99^2}{1-0.99^2}, & \vert\braket{\phi_m|\phi_n}\vert > 0.99, \\
    0, &  \vert\braket{\phi_m|\phi_n}\vert \leq 0.99.
\end{cases}
\end{equation}
The constant $\beta$ can be chosen to increase or decrease the sensitivity to large overlaps.

\subsubsection{Adding and removing basis functions}\label{sec:addremove}
If the procedure at a given time step does not manage to reduce the Rothe error below $\varepsilon_{\Delta t}^2$, up to two basis functions are added. The initial parameters for a single basis function are suggested using the stochastic variational method.~\cite{Bubin_Adamowicz_ECG2013} That is, a large number $K$ of basis functions is produced, with nonlinear parameter $x_i$ following the distribution
\begin{equation}\label{eq:SVM}
    \rho(x_i)=\frac{1}{M} \sum_{n=1}^M \frac{1}{\sqrt{2 \pi\left(\alpha_i^n\right)^2}} \exp \left\{-\frac{\left(x_i-\alpha_i^n\right)^2}{2\left(\alpha_i^n\right)^2}\right\},
\end{equation}
where $\alpha_i^n$ is the $i\ts{th}$ nonlinear parameter of basis function $n$. We use $K=500$ in this work. The Rothe error function is then evaluated for each possible additional basis function, and the basis function that minimizes the Rothe error is added. Subsequently, the nonlinear parameters are reoptimized. If the Rothe error after this reoptimization does not fall below $\varepsilon_{\Delta t}^2$, the procedure is repeated for a second basis function. 
{If the Rothe error is not reduced by at least $1\%$ when adding a new basis function, no basis function is added.}

The addition of basis functions can lead to an unnecessarily large basis set, in the sense that it is possible to get the Rothe error below $\varepsilon_{\Delta t}$ even with fewer basis functions. This can complicate the optimization, and increases the computational cost. To tackle this problem, at least partially, we use the following strategy to remove basis functions, which is carried out at every $10$th time step once the number of Gaussians is greater than the initial number of Gaussians at $t=0$:
\begin{enumerate}
    \item Optimize all nonlinear parameters $\boldsymbol{\alpha}$ and calculate the Rothe error ${\text{r}}^\text{opt.}$. 
    \item For each basis function $k$, $k=1,\dots, N$, calculate the Rothe error $\text{r}_k$ if it were removed from the basis set. Pick the basis function $i=\argmin_k{\text{r}_k}$.
    \item Re-optimize the nonlinear parameters with the basis function $i$ removed, giving rise to an optimized error ${\text{r}}^\text{opt.}_i$.
    \item If ${\text{r}}^\text{opt.}_i<1.01{\text{r}^\text{opt.}}$, remove basis function $i$, as it barely contributes. 
\end{enumerate}
We found that this procedure works well for the system considered, but different systems or different basis functions might require alternative procedures.

\subsection{Calculation of matrix elements}
As mentioned previously in Sec. \ref{sec:Rothe}, Rothe's method requires not only the calculation of the overlap matrix and the Hamiltonian matrix, but also the squared Hamiltonian matrix. The overlap matrix and the Hamiltonian matrix and their derivatives can be calculated in a straightforward way.~\cite{cafiero2001analytical} Matrix elements involving the regularized Coulomb potential $\erf(\mu r)/r$ can be calculated by using the relations
\begin{equation}
    \frac{\erf(\mu r)}{r}=\frac{2}{\sqrt \pi}\int_0^\mu \exp\left(-u^2r^2\right)\ \text{d}u, 
\end{equation}
and 
\begin{equation}
\begin{split}
        &\frac{2}{\sqrt{\pi}}\int_{0}^\mu\frac{1}{\left(1+u^2a\right)^{3/2}}\exp\left(-
        \frac{u^2b}{1+u^2a}\right)\text{d}u\\
        = &\frac{1}{\sqrt{b}}\erf\left(\frac{\sqrt{b}\mu}{\sqrt{a\mu^2+1}}\right).
\end{split}
\end{equation}
Matrix elements of the squared Hamiltonian are straightforwardly derived, except for the term  $\erf^2(\mu r) / r^2$,
which we approximate as 
\begin{equation}
    \frac{\erf^2(\mu r)}{r^2}\approx \frac{1-\exp(\mu^2r^2)}{r^2}+\sum_{k=1}^K c_k\exp(-d_kr^2),
\end{equation}
where $K=10$, and the parameters $c_k, d_k$ were chosen to minimize the difference
\begin{equation}
    \sum_i^N\left(\frac{\erf^2(\mu r_i)}{r_i^2}- \frac{1-\exp(\mu^2r_i^2)}{r_i^2}-\sum_{k=1}^K c_k\exp(-d_kr_i^2)\right)^2,
\end{equation}
for a set of sample points $r_1,\dots,r_N$.
To calculate matrix elements involving $r^{-2}$, we used~\cite{beylkin2005approximation}
\begin{equation}
    \frac{1}{r^2}=2\int^{\infty}_{-\infty}\exp\left(-r^2\text{e}^{2s}+2s\right)\ \text{d}s,
\end{equation}
and
\begin{equation}
\int_{-\infty}^{\infty} \frac{1}{\left(1+\mathrm{e}^{2 u} a\right)^{3/2}} \exp{- \frac{b\mathrm{e}^{2 u}}{1+a\mathrm{e}^{2 u}} }\mathrm{e}^{2 u} \text{d} u=\frac{D\left(\sqrt{\frac{b}{a}}\right)}{\sqrt{b}\sqrt{a}},
\end{equation}
where $D(x)$ is Dawson's function,
\begin{equation}
    D(x)=\frac{\sqrt{\pi}}{2} e^{-x^2} \operatorname{erfi}(x),
\end{equation}
with $\operatorname{erfi}(x)=-\text{i}\operatorname{erf}(\text{i}x)$.

\subsection{The linear Rothe basis}
In Rothe's method, a new set of functions is produced at every time step
\begin{equation}
\mathcal{M}(t)=\{\phi_1(\boldsymbol {\alpha} (t)),\dots,\phi_{M(t)}(\boldsymbol \alpha(t))\}.    
\end{equation}
After a new basis $\mathcal{M}(t)$ has been obtained at each time step, a basis that contains all functions produced in a Rothe propagation can be obtained as 
\begin{equation}
    \mathcal{M}_R=\bigcup_{t\in\{0,\Delta t, 2\Delta t, \dots, T_f\}} \mathcal{M}(t),
\end{equation}
where the union goes over all time steps. This basis set will be very large and hence impractical to use. Instead, we consider the basis set
\begin{equation}
    \mathcal{M}^N_R=\bigcup_{t\in\{0,N\Delta t, 2N\Delta t, \dots\}} \mathcal{M}(t),
\end{equation}
i.e., the basis set produced by only including the functions at every $N$th time step for some $N\in\mathbb{N}$, $N<T_f/(\Delta t)$. \\
From this basis set, we can produce a Hamiltonian matrix $\boldsymbol{H}(t)$ and an overlap matrix $\boldsymbol{S}$, where the time dependence of the Hamiltonian matrix now only stems from the explicit time dependence of the potential, as the basis is now fixed. This basis might be numerically unstable due to (numerical) overcompleteness, so one may use canonical orthogonalization \cite{Szabo&Ostlund} to obtain a smaller, orthonormal basis set, with some cutoff term $s_\varepsilon$ for the smallest eigenvalue of the overlap matrix $\boldsymbol{S}$ considered. We will refer to this fixed basis set as the \textit{linear Rothe basis}. It can potentially be useful on its own, as it is a basis set optimized for HHG calculations written as a linear combination of Gaussians. Its size can also be considered an approximate measure of the compression achieved by Rothe's method. 

\subsection{HHG spectra}

Formally, the HHG spectrum can be calculated as the Fourier transform of the time-dependent dipole velocity\cite{hhg_dpm1,hhg_dpm2}
\begin{equation}
S(\omega)\propto \left| \int_{-\infty}^\infty   \braket{\Psi(t)|\frac{\text{d}z}{\text{d}t}|\Psi(t)}e^{\text i\omega t} \text{d}t \right|^2.
\end{equation}
In practical calculations, the HHG spectrum can only be calculated from a finite-time signal 
\begin{equation}\label{eq:Exact_Fourier}
    S(\omega)\propto \left| \int_{0}^{T_f}   \braket{\Psi(t)|\frac{\text{d}z}{\text{d}t}|\Psi(t)}e^{\text i\omega t} \text{d}t \right|^2,
\end{equation}
which (unlike the infinite-time case) is not necessarily proportional to the Fourier transform of the finite-time dipole moment times $\omega^2$, i.e., 
\begin{equation} \label{eq:z_omega}
z(\omega)\propto \omega^2 \left|  \int_{0}^{T_f} \braket{\Psi(t)|z|\Psi(t)}e^{\text i\omega t} \text{d}t \right|^2.
\end{equation}
Nevertheless, in this paper, we calculate the HHG spectrum as the discrete Fourier transform of the dipole moment multiplied
by the Hann window function.~\cite{Hanning}

\section{Results}\label{sec:Results}

\subsection{Computational details}

To benchmark the performance of Rothe's method with Gaussian wave packets, we compare it to two different methods. 

First, we compare it to a discrete variable representation (DVR) solution of the TDSE, as detailed in Ref.~\citenum{kristiansen2024pseudospectral}. The wavefunction in spherical coordinates is parameterized as 
\begin{equation}
    \Psi(r,\theta,\phi) = \sum_{l=0} ^{l_{\text{max}}} \sum_{m=-l}^l r^{-1}u_{l,m}(r,t)Y_{l,m}(\theta,\phi) \label{spherical_wavefunction_ansatz_full},
\end{equation}
where $Y_{l,m}(\theta,\phi)$ are spherical harmonics, and where $u_{l,m}(r,t)$ vanish at $r=0$ and as $r\to+\infty$. Within the electric-dipole approximation, the quantum number $m$ is conserved when the electric-field vector is parallel to the $z$-axis. As the initial state is taken to be the ground state, a state with $m=0$, it is sufficient only to include spherical harmonics with $m=0$, that is, 
\begin{equation}
    \Psi(r,\theta) = \sum_{l=0} ^{l_{\text{max}}} r^{-1}u_l(r,t)Y_{l,0}(\theta) \label{spherical_wavefunction_ansatz}.
\end{equation} 
The radial functions $u_l(r)$ are resolved on a DVR grid (also referred to as a pseudospectral grid). In particular, we use the Gauss-Legendre-Lobatto (GLL) grid, where the grid points $\{ x_j \}_{j=0}^{N}$ are defined as the zeros of the derivative of the $N$-th degree Legendre polynomial $P_N(x)$,
\begin{equation}
    P_N^\prime(x_j) = 0, \ \ j=1,\cdots,N-1,
\end{equation}
and the boundary points are $x_0 = -1$ and $x_N=1$ (giving a total of $N+1$ grid points).~\cite{boyd2001chebyshev, rescigno2000numerical, bandrauk2011quantum, cinal2020highly} The GLL grid points $\{ x_j \}_{j=0}^{N}$  are defined on the interval $[-1,1]$ and are mapped to the radial coordinate $r \in [0, r_{\text{max}}]$ ($r_{\text{max}}$ is a finite chosen cut-off) by a linear mapping,
\begin{equation}
    r(x_j) = \frac{r_{\text{max}}}{2}(x_j+1).
\end{equation}
Insertion of the ansatz~(\ref{spherical_wavefunction_ansatz}) into the TDSE yields a set of coupled equations of motion for the radial functions. In the DVR method, these equations are approximated using collocation, equivalently all matrix elements are approximated with the underlying quadrature of the DVR grid. The resulting spatially discrete Schrödinger equation is propagated using the Crank-Nicolson method.
The convergence parameters used in the implementation described in Ref.~\citenum{kristiansen2024pseudospectral} are set to $r_{\text{tol}} = 10^{-10}$ and $a_{\text{tol}}=0.0$
in the \emph{biconjugate gradients stabilized} 
(Bi-CGSTAB)~\cite{van_der_vorst_bi-cgstab_1992} algorithm.

For all simulations, we have used $r_{\text{max}} = 600\,\text{a.u.}$ and a polynomial order of $N=1200$. 
For grid methods, it is customary to use an absorber or a complex absorbing potential to avoid unphysical reflections at the grid boundary. However, with the chosen $r_{\text{max}}$ the (radial) wavefunction is effectively zero at the grid boundary for all the simulations (see Figs.~\ref{fig:hhg_conv_003}--\ref{fig:hhg_conv_012}) and, therefore, no absorber has been used.
The initial state for the propagation is the ground state found by solving the time-independent Schr{\"o}dinger equation by diagonalization, see Ref.~\citenum{kristiansen2024pseudospectral} for details.
For all the presented examples, the simulations are converged with respect to the angular momentum $l_{\text{max}}$, as can be seen in Figs.~\ref{fig:hhg_conv_003}--\ref{fig:hhg_conv_012}. 

Second, we compare with time evolution in a fixed Gaussian basis, namely the 6-aug-cc-pVTZ+8K basis set described in Ref.~\citenum{Coccia}. This basis set consists of $140$ basis functions. It should be noted, however, that the basis set can be adapted in such a way that only $48$ basis functions contribute due to the cylindrical symmetry. For the fixed Gaussian basis, we use the same heuristic lifetime model
as \citeauthor{Coccia},~\cite{Coccia} i.e., the orbital energies are replaced by complex energies as
\begin{align}
    &E_k \leftarrow E_k- \text{i} \Gamma_k/2, \\
    &\Gamma_k=
    \begin{cases}
          0 &  E_k <0\\
          \sqrt{2E_k} / d_0 &  0 \leq E_k \leq 3.17U_p\\
          \sqrt{2E_k} / d_1 & E_k > 3.17U_p\\
    \end{cases},
\end{align}
where $d_0=50$, $d_1=0.1$, and
\begin{equation}\label{eq:Ponderomotive_energy}
    U_p=\frac{E_0^2}{4\omega^2},
\end{equation}
is the ponderomotive energy of the electron.

Propagation is carried out using the Crank-Nicolson scheme for both the grid calculation and time evolution in the Gaussian basis, as Rothe's method is an approximation to exact Crank-Nicolson propagation. In all simulations, unless otherwise stated, the duration of the driving laser pulse is three optical cycles $N_c=3$ with a carrier frequency $\omega=0.057\,\text{Ha}$, roughly corresponding to a wavelength of $800\,\text{nm}$. The time step is $\Delta t=0.2\,\text{a.u.}$, both for the DVR and the fixed-Gaussian simulations.

We consider three laser intensities, two below and one above the barrier-suppression ionization threshold for the hydrogen atom ($1.137\times 10^{14}\,\text{W/cm}^2$ corresponding to $E_0=0.0625\,\text{a.u.}$),~\cite{Barrier} namely $E_0=0.03\,\text{a.u.}$, $E_0=0.06\,\text{a.u.}$, and $E_0=0.12\,\text{a.u.}$, corresponding to laser peak intensities of $3.16\times 10^{13}\,\text{W/cm}^2$, $1.26\times 10^{14}\,\text{W/cm}^2$ and $5.05\times 10^{14}\,\text{W/cm}^2$, respectively.

For the DVR reference simulations, we use $l_{max}=20$ for $E_0=0.03\,\text{a.u.}$, $l_{max}=30$ for $E_0=0.06\,\text{a.u.}$, and $l_{max}=70$ for $E_0=0.12\,\text{a.u.}$.
For the Rothe propagations, the ground state is approximated using $25$ Gaussians, which remain frozen at all times, and $25$ additional Gaussians centered around the origin are included at $t=0$ to avoid adding Gaussians at the very beginning while keeping the Rothe error small. 
The constant $\beta$ of the penalty term, eq.~\eqref{eq:penalty}, is set to $10^{-13}$ for $E_0=0.03\,\text{a.u.}$ and to $2\times 10^{-12}$ for the higher intensities.
For the field strength $E_0=0.12\,\text{a.u.}$, we do not remove any Gaussians, as the removal procedure described in Sec. \ref{sec:addremove} only led to increased run times for the large Rothe errors considered.

To quantify the agreement of the computed HHG spectra with the DVR reference, we use the three global descriptors proposed by \citeauthor{Morassut}~\cite{Morassut}
The local descriptor $\delta_i$ for the deviation in peak height from the DVR is defined as
\begin{equation}
    \delta_i=\log_{10}(h_i)-\log_{10}(h_i^{\text {DVR}})
\end{equation}
where $h_i$ stands for the harmonic peak value at the $i$'th harmonic obtained using an approximate method and $h_i^{\text {DVR}}$ is the peak height obtained with the reference DVR method. We then consider the following global descriptors
\begin{align}
\Delta_N &=\sum_{i=1}^N \frac{|\delta_i|}{N}\\
 \Upsilon_N &=\sum_{i=1}^N \frac{|\delta_i|^2}{N},
\end{align}
where the summations run over the peaks in the HHG spectrum, i.e., all odd integers. As the time step chosen by us did not yield exact integer values, we used the values closest to odd integers.

In addition, we also consider the descriptor
\begin{align}
    \hat{D}_{\text{corr},N}=\sum_{i=1}^N \Big(&\log_{10}(h_i^{\text {DVR}})\log_{10}(h_i^{\text {DVR}}) \nonumber \\
    -&\log_{10}(h_i^{\text {DVR}})\log_{10}(h_i) \Big),
\end{align}
where the summation runs over \textit{all} frequencies available, not just the peaks. We follow Ref.~\citenum{Morassut} and consider those values for $N$ that correspond to once or twice the classical cutoff obtained from the three-step model $N_{3SM}$,
\begin{equation}
    N_{3SM}=\frac{I_p+3.17U_p}{\omega},
\end{equation}
where $U_p$ is the ponderomotive energy, eq. \eqref{eq:Ponderomotive_energy}, and $I_p=0.5\,\text{Ha}$ is the ionization energy of the hydrogen atom.

\subsection{Regularization of the Coulomb potential}

We start by investigating the effect of regularization on the HHG spectra using DVR simulations with the real-space mesh fine enough to resolve the features of the wave function and the potential near the origin. For reference, the potentials close to the origin are plotted in Fig.~\ref{fig:erf_at_grid_vals} in the appendix.
\begin{figure}[H]
    \centering
    \includegraphics[width=0.5\textwidth]{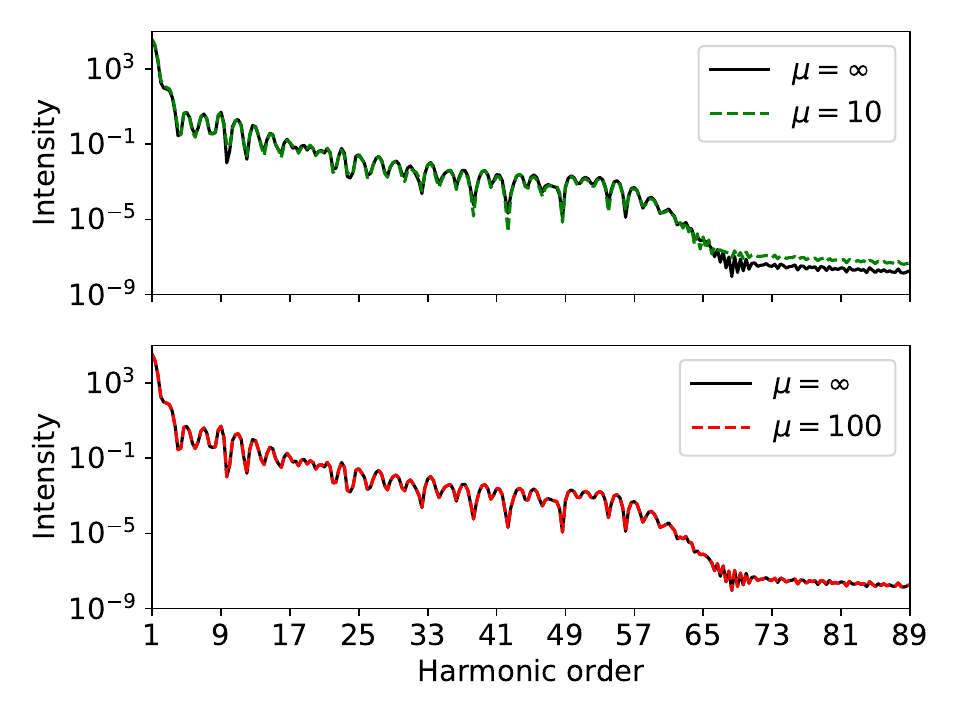}
    \caption{HHG spectra produced with DVR simulations using the Hamiltonian in eq. \eqref{eq:Hamiltonian} with $E_0=0.12\,\text{a.u.}$, $\mu \in \{10,100,\infty\}$, $N_c=3$, and $\omega=0.057\,\text{Ha}$.}
    \label{fig:Coulomb_erfCoulomb_comparison}
\end{figure}

Figure \ref{fig:Coulomb_erfCoulomb_comparison} shows the HHG spectrum for $\mu=10$ and $\mu=100$ compared with the exact Coulomb potential with electric-field strength $E_0=0.12\,\text{a.u.}$
Although the spectrum for $\mu=10$ is very similar to the spectrum obtained using the Coulomb potential, there are visible deviations.
With $\mu=100$, however, the regularized potential and the Coulomb potential yield essentially identical spectra, justifying the use of the regularized Coulomb potential. 

\subsection{Assessing Rothe's method}

\subsubsection{HHG spectra}
\begin{figure}[h]
    \centering
    \includegraphics[width=0.5\textwidth]{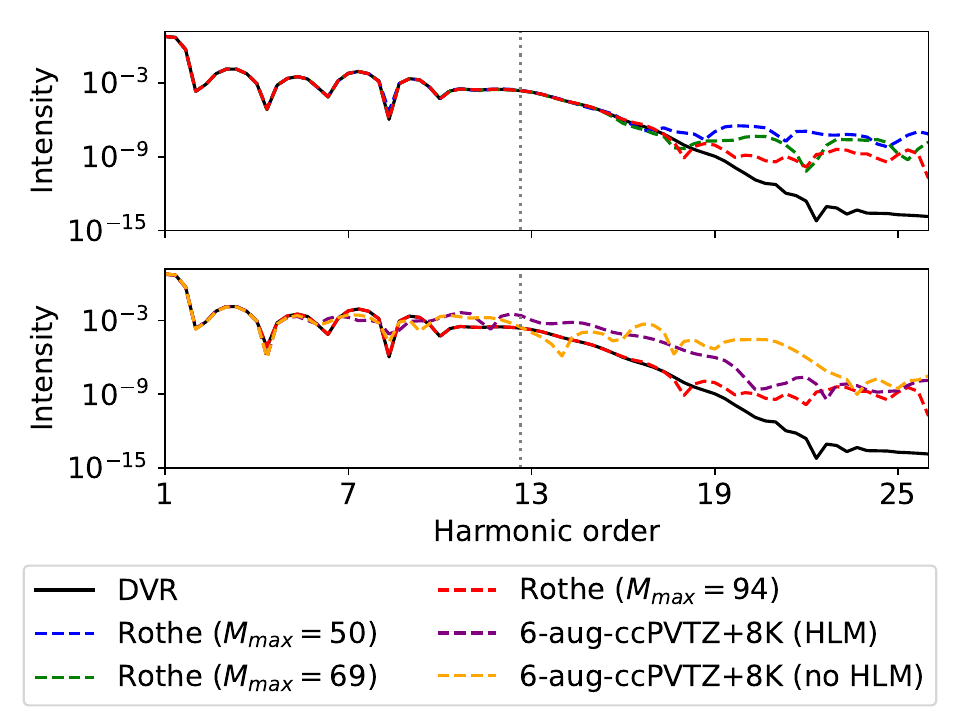}
    \caption{HHG spectra for $E_0=0.03\,\text{a.u.}$, $\omega=0.057\,\text{Ha}$, and $N_c=3$. Top: HHG spectra obtained using Rothe's method with different error thresholds and thus different maximal numbers of Gaussians $M_\text{max}$, compared with the DVR reference simulation. Bottom: Comparing the HHG plot obtained from the Rothe method using $M_\text{max}=94$ Gaussians with the fixed 6-aug-ccPVTZ+8K basis set with and without a heuristic lifetime model (HLM). The cumulative Rothe error is $\varepsilon_{\text{tot}}=0.0027$ for $M_\text{max}=94$,
$\varepsilon_{\text{tot}}=0.0084$ for $M_\text{max}=69$, and $\varepsilon_{\text{tot}}=0.028$ for $M_\text{max}=50$.}
    \label{fig:Rothe_E003}
\end{figure}

Figure \ref{fig:Rothe_E003} shows the HHG spectrum produced using Rothe's method with different numbers of Gaussians for $E_0=0.03$ and compares it with the reference methods described above. In the fixed basis, we consider time evolution with and without a heuristic lifetime model (HLM). We observe that using up to $94$ Gaussians ($\varepsilon_{\text{tot}}=0.0027$), the reference HHG spectrum is essentially reproduced exactly, except for slight noise for large harmonics, and the obtained spectrum resembles the reference spectrum more closely than using the 6-aug-cc-pVTZ+8K basis, both with and without HLM. Reducing the number of Gaussians (and thereby increasing the Rothe error), we observe that the amount of peaks, their intensity and their position are still very well reproduced. We observe a decrease in quality for high harmonic orders as the number of Gaussians is decreased. This is also apparent from the values of the global descriptors shown in table \ref{tab:descriptors_003}, where we see that the HHG spectra are well reproduced up to the three-step model threshold, as can be seen from the small values in $\Delta_{13}$, $\Upsilon_{13}$ and $D_{13}$ even for $M=50$, improving those of the 6-aug-cc-pVTZ+8K basis. We observe that global descriptors improve significantly by increasing the number of Gaussians, and the improvement for $\Delta_{13}$, $\Upsilon_{13}$ and $D_{13}$ indicates that the spectrum is converging to the DVR calculation before the three-step model threshold, while the improvement in $\Delta_{25}$, $\Upsilon_{25}$ and $D_{25}$ indicates that more Gaussians also improve the HHG spectrum after the classical cutoff.
Visual comparison with Fig.~\ref{fig:hhg_conv_003} in the appendix also shows that using even just $50$ Gaussians already yield results as good as $l_{max}=10$ before the classical cutoff, indicating that the flexible Gaussian ansatz is effectively able to reproduce higher-order angular momentum states.

\begin{figure}[h]
    \centering
    \includegraphics[width=0.5\textwidth]{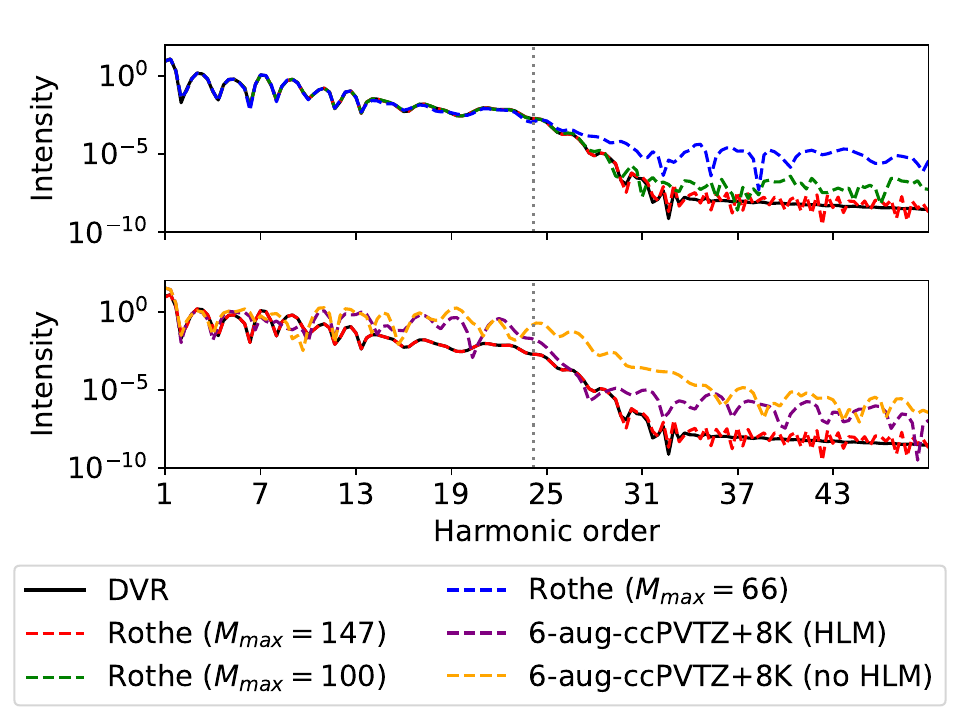}
    \caption{HHG spectra for $E_0=0.06\,\text{a.u.}$, $\omega=0.057\,\text{Ha}$, and $N_c=3$. Top: 3 HHG spectra obtained using Rothe's method using different error thresholds and thus getting different maximal numbers of Gaussians $M_\text{max}$, compared to the DVR calculation. Bottom: Comparing the HHG plot obtained by Rothe's method using$M_\text{max}=147$ Gaussians to the 6-aug-cc-pVTZ+8K basis set with a heuristic lifetime model (HLM). The cumulative Rothe error is $\varepsilon_{\text{tot}}=0.013$ for $M_\text{max}=147$,
$\varepsilon_{\text{tot}}=0.085$ for $M_\text{max}=100$, and $\varepsilon_{\text{tot}}=0.39$ for $M_\text{max}=66$.}
    \label{fig:Rothe_E006}
\end{figure}

\begin{figure}[h]
    \centering
    \includegraphics[width=0.5\textwidth]{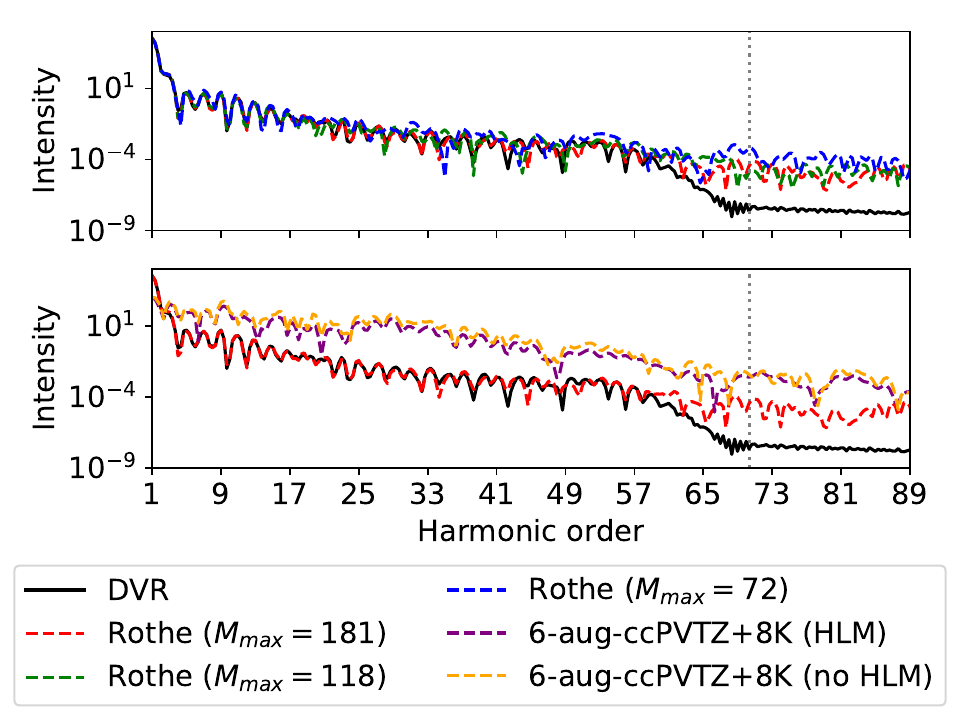}
    \caption{HHG spectra for $E_0=0.12\,\text{a.u.}$, $\omega=0.057\,\text{Ha}$, and $N_c=3$. Top: 3 HHG spectra obtained using Rothe's method using different error thresholds and thus getting different maximal numbers of Gaussians $M_\text{max}$, compared to the DVR calculation. Bottom: Comparing the HHG plot obtained by Rothe's method using $M_\text{max}=150$ Gaussians to the 6-aug-cc-pVTZ+8K basis set with a heuristic lifetime model (HLM). The cumulative Rothe error is $\varepsilon_{\text{tot}}=1.49$ for $M_\text{max}=181$, 
$\varepsilon_{\text{tot}}=2.52$ for $M_\text{max}=118$, and $\varepsilon_{\text{tot}}=3.82$ for $M_\text{max}=72$.}
    \label{fig:Rothe_E012}
\end{figure}

\begin{table}[h]
    \centering
    \begin{NiceTabular}{c|cc:cc:cc}
        Method & $\Delta_{13}$ & $\Delta_{25}$ & $\Upsilon_{13}$ & $\Upsilon_{25}$ & $\hat{D}_{\text{corr},13}$ & $\hat{D}_{\text{corr},25}$\\\hline
        Rothe (M=50) & 0.0037 &0.73   & 0.000078&3.7   &4.7  & 1200 \\
        Rothe (M=69) &0.0026  & 0.63  & 0.000020 &2.8  &0.68  & 990 \\
        Rothe (M=94) &0.0020  &0.50   & 0.000012  &2.0 &0.52  & 790 \\\hdottedline
        FGB (HLM)    & 0.23  & 0.91 &0.17   &3.1       &35  & 1200 \\
        FGB (no HLM) &0.14   &1.09    &0.064           &5.2  &21&1600 \\
    \end{NiceTabular}
    \caption{Global descriptors $\Delta_N$, $\Upsilon_N$ and $\hat{D}_N$ for $E_0=0.03\,\text{a.u.}$, $\omega=0.057\,\text{Ha}$, and $N_c=3$ using the DVR simulation as reference. $N=13$ corresponds to the classical cutoff, while $N=25$ is twice that. $M$ refers to the maximal number of Gaussians used, and FGB to the fixed Gaussian basis 6-aug-cc-pVTZ+8K. }
    \label{tab:descriptors_003}
\end{table}
Figure \ref{fig:Rothe_E006} shows the HHG spectrum produced using Rothe's method with different numbers of Gaussians for $E_0=0.06\,\text{Ha}$ and compares it to the previously described reference methods, with table \ref{tab:descriptors_006} containing the corresponding global descriptors. Here, we obtain a spectrum that closely resembles the DVR calculation by using up to $M_\text{max}=147$ Gaussians both before and after the classical cutoff, i.e., the HHG spectrum is quantitatively correct everywhere. Using $M_\text{max}=100$ Gaussians, we obtain a quantitatively correct spectrum before the classical cutoff, while the spectrum after the classical cutoff is qualitatively between the 6-aug-cc-pVTZ+8K basis and the reference. Even using only $M_\text{max}=66$ Gaussians gives quantitatively correct results before the classical cutoff and shows some reduction in intensity after the cutoff. This again shows that relatively few Gaussians are needed to calculate the HHG spectrum before the classical cutoff, while the amount of Gaussians needs to be more than doubled to obtain a quantitatively correct spectrum also after the classical cutoff. While the 6-aug-cc-pVTZ+8K basis set performs much worse for $E_0=0.06\,\text{a.u.}$ than for $E_0=0.03\,\text{a.u.}$, we observe that our approach performs similarly well for both field strengths compared to the DVR calculation. Indeed, for $E_0=0.06\,\text{a.u.}$, the DVR calculation is almost exactly reproduced using $M_\text{max}=147$ Gaussians. We also observe that even though we get a Rothe error of $1.3\%$ for $M_\text{max}=147$ Gaussians, which is approximately five times higher than the Rothe error we get for $M_\text{max}=94$ for $E_0=0.03\,\text{a.u.}$, the spectrum is much closer to its respective reference spectrum. This shows that larger Rothe errors for stronger fields do not necessarily indicate that the results become qualitatively worse.
Figure \ref{fig:hhg_conv_006} in the appendix also shows that $l_{max}=20$ is required in the DVR simulation to obtain a quantitatively correct spectrum before the classical cutoff. The same quality is obtained with with only $66$ Gaussians in the Rothe propagation.
\begin{table}[H]
    \centering
    \begin{NiceTabular}{c|cc:cc:cc}
        Method & $\Delta_{25}$     & $\Delta_{49}$ & $\Upsilon_{25}$ & $\Upsilon_{49}$ & $\hat{D}_{\text{corr},25}$ & $\hat{D}_{\text{corr},49}$\\\hline
        Rothe (M=66)  & 0.023  &0.64  &  0.0016&1.87  &-4.2  & 1400 \\
        Rothe (M=100) &0.0062  & 0.23 & 0.00013 &0.29  &0.38  & 490 \\
        Rothe (M=147) &0.0033  &0.074  &0.000070 &0.033  &0.20  & 47 \\\hdottedline
        FGB (HLM)     & 0.43 & 0.61 &0.49  &0.99  &100  & 970 \\
        FGB (no HLM)  &0.55  &0.95  &0.87  &2.33  &130  &1600 \\
    \end{NiceTabular}
    \caption{Global descriptors $\Delta_N$, $\Upsilon_N$ and $\hat{D}_N$ for $E_0=0.06\,\text{a.u.}$, $\omega=0.057\,\text{Ha}$, and $N_c=3$, where a DVR calculation is used as a reference. $N=25$ corresponds to the classical cutoff, while $N=49$ is twice that. $M$ refers to the maximal number of Gaussians used, and FGB to the fixed Gaussian basis 6-aug-cc-pVTZ+8K.}
    \label{tab:descriptors_006}
\end{table}
Finally, we consider the field strength $E_0=0.12\,\text{a.u.}$ The obtained HHG spectra and global descriptors are shown in Fig.~\ref{fig:Rothe_E012} and table \ref{tab:descriptors_012}, respectively. While the 6-aug-cc-pVTZ+8K basis gives qualitatively correct results for $E_0=0.03\,\text{a.u.}$ and $E_0=0.06\,\text{a.u.}$, the HHG spectra for $E_0=0.12\,\text{a.u.}$ are qualitatively wrong. Rothe's method yields spectra that are quantitatively correct up to approximately the $25$th harmonic using $M_\text{max}=72$ basis functions, and quantitative agreement up to approximately the $60$th harmonic can be obtained using $M_\text{max}=181$ Gaussians. While the intensity is orders of magnitudes too high after the classical cutoff using Rothe's method, we observe an improvement in the quality of the spectra as the number of Gaussians is increased, which indicates that improved agreement with the reference calculation is possible by increasing the number of Gaussians. 
Comparison with figure \ref{fig:hhg_conv_012} in the appendix shows that $181$ Gaussians produces a spectrum which is roughly between $l_{max}=30$ and $l_{max}=40$ up to harmonic order $\sim 60$, again showing that Gaussians can reproduce the effects that are attributed to the inclusion of high angular momenta.
\begin{table}[h]
    \begin{NiceTabular}{c|cc:cc:cc}
        Method & $\Delta_{71}$ & $\Delta_{141}$ & $\Upsilon_{71}$ & $\Upsilon_{141}$ & $\hat{D}_{\text{corr},71}$ & $\hat{D}_{\text{corr},141}$\\\hline
        Rothe (M=72) &0.36 &1.1  &0.92 &3.9  & 720 & 7100 \\
        Rothe (M=118)&0.26 &0.85  &0.45 &2.4  & 550  & 5600 \\
        Rothe (M=181)&0.22 &0.76 &0.49  &2.1  &420  & 5000\\\hdottedline
        FGB (HLM)    & 1.2 &1.6 &3.4   &6.1  & 1700  & 8700\\
        FGB (no HLM) &1.3  &1.8  &4.2   &7.0  & 2000 & 9300\\
    \end{NiceTabular}
    \caption{Global descriptors $\Delta_N$, $\Upsilon_N$ and $\hat{D}_N$ for $E_0=0.12\,\text{a.u.}$, $\omega=0.057\,\text{Ha}$, and $N_c=3$, where a DVR calculation is used as a reference. $N=71$ corresponds to the classical cutoff, while $N=141$ is twice that. $M$ refers to the maximal number of Gaussians used, and FGB to the fixed Gaussian basis 6-aug-cc-pVTZ+8K. }
    \label{tab:descriptors_012}
\end{table}

\subsubsection{The linear Rothe basis}
\begin{figure}[H]
    \centering
    \includegraphics[width=0.5\textwidth]{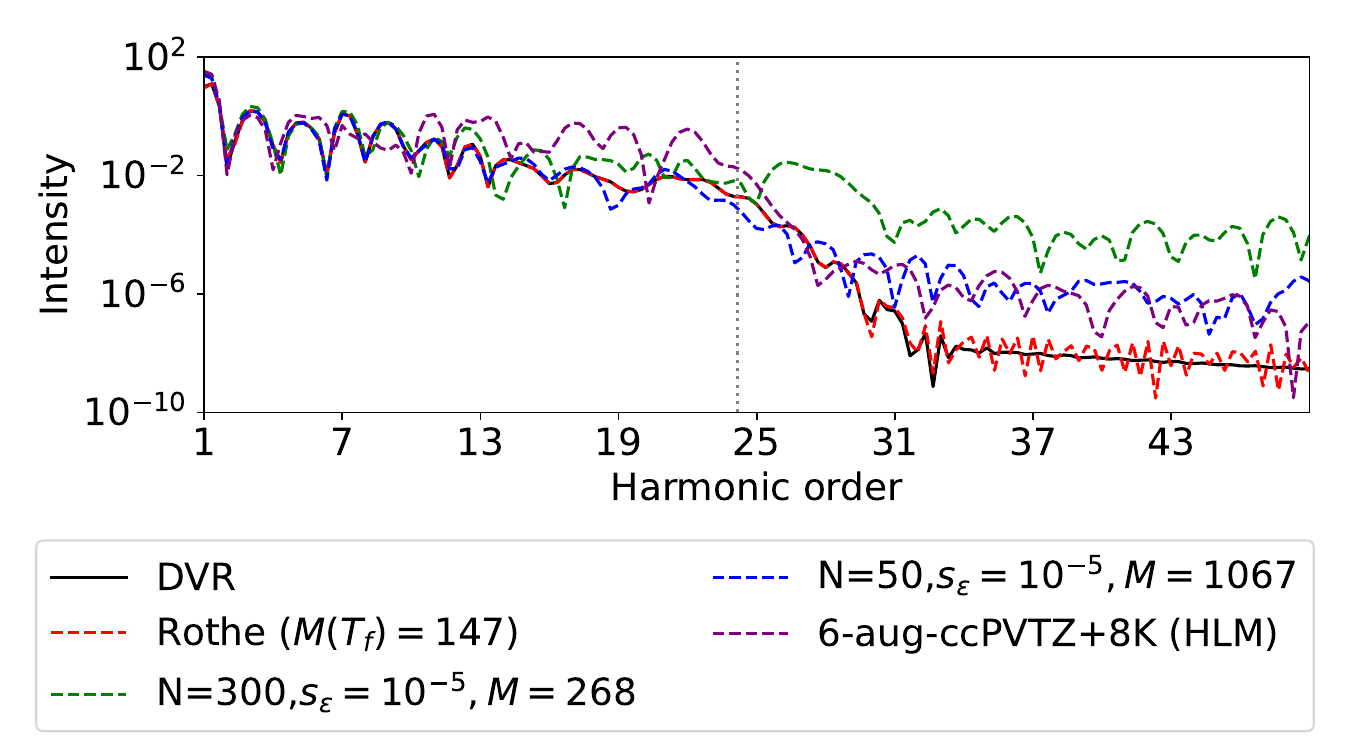}
    \caption{Comparison of HHG spectra for $E_0=0.06\,\text{a.u.}$, $\omega=0.057\,\text{Ha}$, and $N_c=3$ obtained from simulations with DVR, the 6-aug-cc-pVTZ+8K basis, Rothe's method with $M_\text{max}=147$ Gaussians, and the linear Rothe basis generated from Rothe simulations with $N=300$, $s_\varepsilon=10^{-5}$, and $M=268$, and $N=50$, $s_\varepsilon=10^{-5}$, and $M=1067$ basis functions. The heuristic lifetime model is used for the linear Rothe bases and for the 6-aug-cc-pVTZ+8K basis.}
    \label{fig:Rothe_bas}
\end{figure}

Finally, we consider the dimension of the linear Rothe basis needed to reproduce a qualitatively correct HHG spectrum. For these fixed basis sets, the heuristic lifetime model is used. The linear Rothe basis produced by sampling every $300$th time step with the pseudoinverse cutoff set to $s_\varepsilon=10^{-5}$, giving a total of $M=268$ basis functions, is shown in Fig.~\ref{fig:Rothe_bas}. Sampling every $50$th time step with a cutoff of $s_\varepsilon=10^{-5}$ leads to $M=1067$ basis functions. We see that using $M=268$ basis functions, the spectrum is well reproduced up to the classical cutoff with greater errors beyond. Better results are achieved using $M=1067$ basis functions, which is a good approximation to the full Rothe calculation before the classical cutoff and on par with the 6-aug-cc-pVTZ+8K basis calculation after. We conclude that the time evolution of the nonlinear parameters makes it possible to produce a fixed basis set which can be used to produce high-quality HHG spectra, and that our choice of shifted-center Gaussian basis functions can make for an efficient basis set for time-dependent calculations. However, our approach is very simple and the resulting basis with $M=1067$ is extremely large. We believe that instead fitting basis functions to the space of functions spanned by the Rothe functions at all time steps, reminiscent of the fitting procedure performed by \citeauthor{Wozniak},~\cite{Wozniak} can substantially reduce the number of basis functions needed to achieve accurate spectra.

\section{Concluding remarks}\label{sec:Conclusion}
We have shown that Rothe's method, which recasts time evolution as an optimization problem, can be used to propagate the hydrogen wave function in a strong laser field using an adaptive basis of shifted, complex Gaussian functions with variable widths, positions, and momenta determined at each time step. We studied the applicability of the approach by examining the HHG spectra obtained and found that Rothe's method with shifted, complex Gaussians can generate quantitatively correct HHG spectra with only 60-150 basis functions for field strengths up to $\sim 5\times 10^{14}\,\text{W/cm}^2$ without the need for complex absorbing potentials, mask functions, or heuristic lifetime models. In particular, for strong fields, we discovered that by allowing for a variation of the nonlinear parameters, the number of necessary basis functions for time-dependent calculations can be reduced from hundreds to only dozens, while the quality of the resulting HHG spectra remains approximately the same. Rothe's method paves the way for studying systems in intense laser fields when grid calculations become infeasible and for which no efficient fixed basis sets exist, as Rothe's method itself builds a suitable, minimal basis at every time step. The resulting wave functions are extremely compact at every time step, and using time-dependent Gaussians, the exponential scaling in memory usage that arises with grid methods is alleviated. We conclude that Rothe's method is a viable alternative to time-dependent variation principles and enables us to model time-dependent phenomena with high accuracy.

Rothe's method can be extended to mean-field methods such as real-time time-dependent Hartree-Fock theory, or it can be adapted to be applicable in an MCTDHF context.~\cite{Zanghellini} Furthermore, it is conceptually straightforward to use explicitly correlated Gaussians\cite{Bubin_Adamowicz_ECG2013,Mitroy_Gaussian2013} to consider the time evolution of correlated N-particle systems, including atoms and molecules. Thus, Rothe's method paves the way for highly accurate descriptions of the time evolution of small molecules without invoking the Born-Oppenheimer approximation at any stage. This can be done without constructing a basis for anything but the initial ground state beforehand\cite{Adamowicz_Kvaal_Lasser_Pedersen} and without being restricted to few-dimensional model systems.~\cite{Non_BO_H2plus} Although it breaks cylindrical symmetry and, therefore, requires a modified algorithm for addition and removal of basis functions, going beyond the electric-dipole approximation is also viable, as it only leads to different matrix elements for the matter-field interaction.~\cite{aurbakken_transient_2024}

Since Rothe's method does not require the solution of differential equations, a very flexible basis can in principle be chosen. In particular, it should be possible to represent the wave function on an adaptive grid, using either a multiresolution scheme\cite{LORIN_multiresolution} or a tensor network representation.~\cite{jolly2023tensorized,Soley2022} Another interesting possibility would be a neural-network representation that explicitly constructs a spatial wave function, such as PauliNet\cite{PauliNet} or FermiNet.~\cite{FermiNet}

\section*{Data availability}
The data that supports the finding of this study is available on Zenodo, see Ref.~\citenum{Rothe_code}.
\section*{Author Declaration}
The authors have no conflicts to disclose.

\section*{Acknowledgements}
We thank Profs.~Caroline Lasser and Ludwik Adamowicz for helpful discussions. The work was supported by the Research Council of Norway through its Centres of Excellence funding scheme, Project No. 262695.

\bibliography{main}
\appendix
\counterwithin{figure}{section}
\section{Matrix elements}
Here, we give the formulas used for calculating matrix elements.
Defining
\begin{equation}
        \phi_k(x,y,z)=\text{e}^{-(a_k^2+\text{i}b_k)(x^2+y^2+(z-\mu_k)^2)+\text{i}q_k(z-\mu_k)},
\end{equation}
and 
\begin{align}
    c_k&=a_k^2+\text{i}b_k,\\
    M_k&=2c_k\mu_k+\text{i}q_k,\\
    c_{lk}&=c_l^*+c_k,\\
    {y}_{lk}&=2c_l^*\mu_l+2c_k\mu_k+\text{i}(q_k-q_l),\\
    \gamma_{lk}&=\text{i}(\mu_l^Tq_l-\mu_kq_k)-c_l^*\mu_l\mu_l-c_k\mu_k^T\mu_k,
\end{align}
we find the overlap as 
\begin{equation}
    \braket{\phi_l|\phi_k}=\frac{\pi^{3/2}}{\sqrt{{c}_{lk}}^3}\exp\left(\frac{1}{4{c}_{lk}}{y}_{lk}^2+\gamma_{lk}\right).
\end{equation}
(As $c_{lk}$ is complex, note that $\sqrt{{c}_{lk}}^3\neq \sqrt{{c}_{lk}^3}$.)
Expectation values involving the regularized Coulomb potential can be calculated as 
\begin{align}
\frac{\bra{\phi_l}\erf(\mu r)/r\ket{\phi_k}}{2} &=\braket{\phi_l|\phi_k}\frac{{c}_{lk}}{ y_{lk}}\nonumber \\
 &\times \erf\left(\frac{0.5\sqrt{y_{lk}^2}\mu}{\sqrt{\mu^2{c}_{lk}+{c}_{lk}^2}}\right),
\end{align}
while expectation values involving the squared Coulomb potential read
\begin{equation}
\bra{\phi_l}1/r^2\ket{\phi_k}=4\braket{\phi_l|\phi_k}D\left(0.5\sqrt{\frac{  y_{lk}^2}{{c}_{lk}}}\right)\frac{\left(\sqrt{c_{lk}}\right)^{3}}{\sqrt{ y_{lk}^2}}.
\end{equation}
Matrix elements involving the Laplacian $\nabla^2$, the external potential $z$, the Laplacian squared $\nabla^4$ and the cross terms between the Laplacian and the potentials ($\nabla^2\cdot z$, $z\cdot \nabla^2$,  $\nabla^2 \cdot \erf(\mu r)/r$ and $\erf(\mu r)/r \cdot \nabla^2$) can all be calculated as linear combination of derivatives of the overlap or the potential. For example, the expectation value of the Laplacian reads
\begin{align}
    &\bra{\phi_l}\nabla^2\ket{\phi_k}= \nonumber  \\
    &\left(\!-4c_k^2\frac{\partial}{\partial c_{lk}}-4c_k{M_{k}}\frac{\partial}{\partial {{y}_{lk}}}+{M_{k}^2}-6c_k\!\!\right)\nonumber \\
    &\times \braket{\phi_l|\phi_k},
\end{align}
and the cross term $(\erf(\mu r)/r) \nabla^2 $ becomes
\begin{align}
    &\bra{\phi_l}(\erf(\mu r)/r) \nabla^2\ket{\phi_k}=\nonumber \\
    &\left(-4c_k^2\frac{\partial}{\partial c_{lk}}-4c_k{M_{k}}\frac{\partial}{\partial {{y}_{lk}}}+{M_{k}^2}-6c_k\right) \nonumber \\
    &\times \bra{\phi_l}\erf(\mu r)/r\ket{\phi_k}.
\end{align}
Some of the terms considered here are undefined for $y_{lk}=0$. In particular, the basis functions representing the ground state have that $y_{lk}=0$. This is however not a problem, as the Taylor expansions are defined also for $y_{lk}=0$ (letting $y_{lk}/y_{lk}$=1 even if $y_{lk}=0$). We used a fourth-order Taylor expansion for expectation around $y_{lk}=0$ for $y_{lk}<10^{-3}$.

\section{DVR}

Within the electric-dipole approximation, the parametrization of the wavefunction is given by eq.~\eqref{spherical_wavefunction_ansatz}.
To analyse the wavefunction at a given time $t$, we define the angular distribution as
\begin{equation}
    P_l(\Psi(\mathbf{r},t)) = \int_0^\infty |u_l(r,t)|^2 dr,
\end{equation}
and the radial distribution as
\begin{equation}
    R(r,t) = \sum_{l=0}^{l_{\text{max}}} |u_l(r,t)|^2.
\end{equation}
\begin{figure*}
    \centering  
    \includegraphics[width=\textwidth]{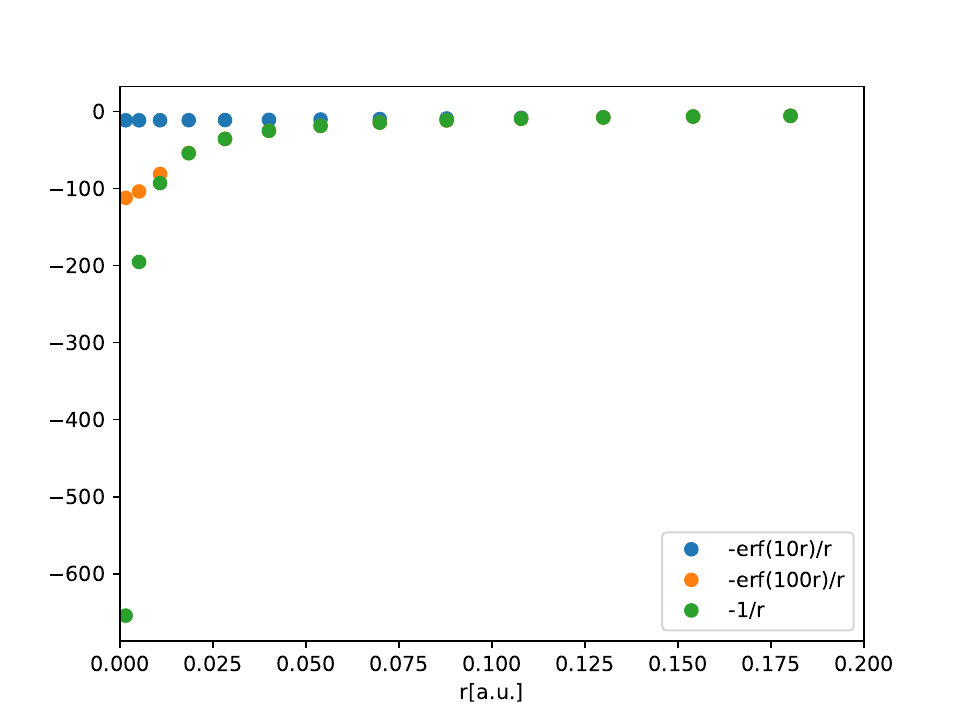}
    \caption{Values of $-\erf\left({\mu r}\right)/r$ for $\mu\in\{10,100\}$ and $-1/r$ at the $13$ first inner grid points when using $r_{\text{max}}=600\,\text{Bohr}$ and $N=1200$ as polynomial order.}
    \label{fig:erf_at_grid_vals}
\end{figure*}

\begin{figure*}
\begin{subfigure}[b]{0.6\textwidth}
\centering
   \includegraphics[scale=0.6]{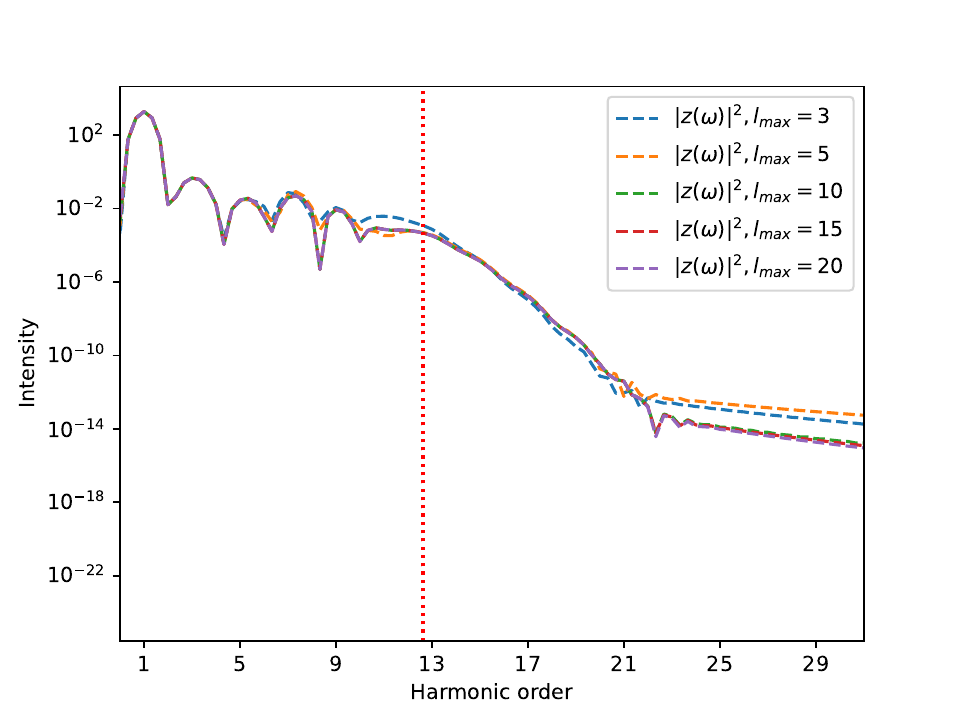}
\end{subfigure}
\begin{subfigure}[b]{0.6\textwidth}
   \includegraphics[scale=0.6]{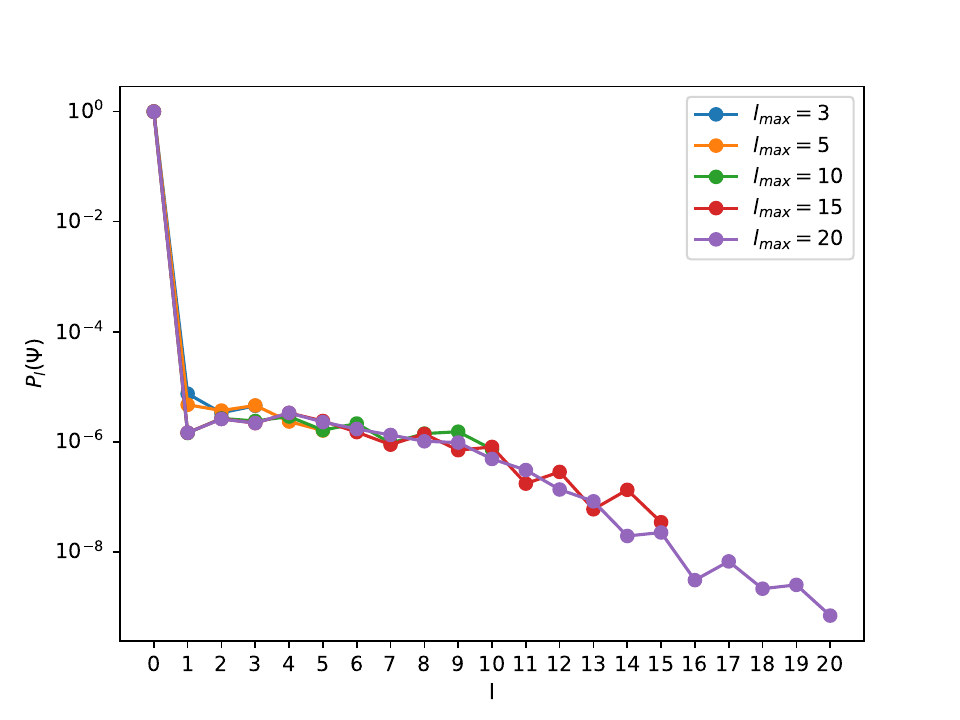}
\end{subfigure}
\begin{subfigure}[b]{0.6\textwidth}
   \includegraphics[scale=0.6]{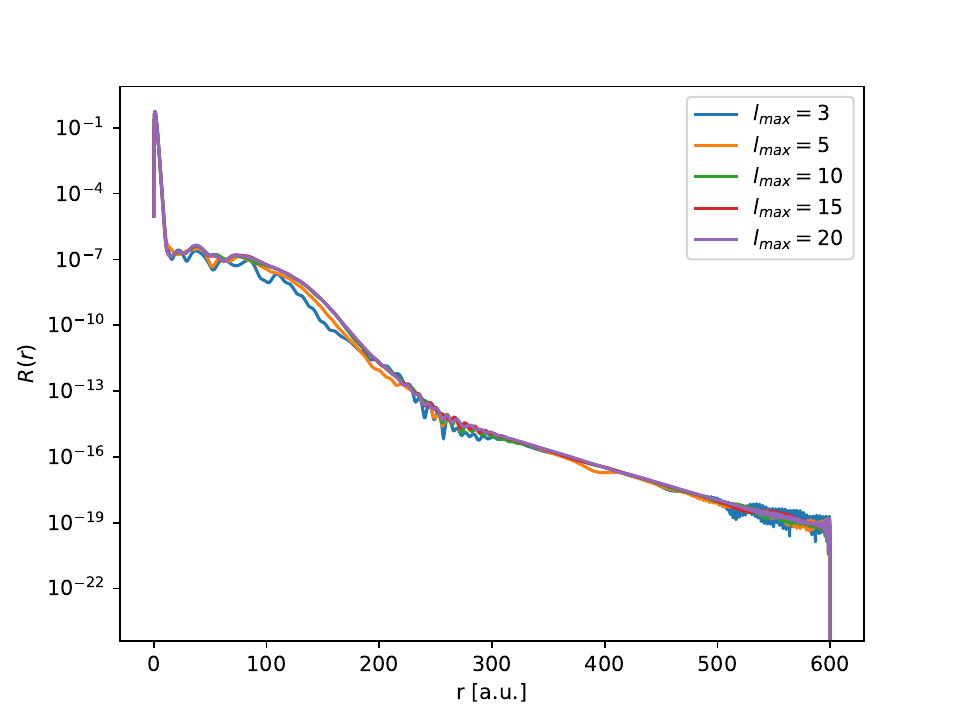}
\end{subfigure}
    \caption[Description]{HHG spectrum, angular momentum and radial distribution at $t=T_f$ for $E_0 = 0.03\,\text{a.u.}$ using $r_{\text{max}}=600\,\text{Bohr}$ and $N=1200$.}
\label{fig:hhg_conv_003} 
\end{figure*}

\begin{figure*}
\centering
\begin{subfigure}[b]{0.6\textwidth}
   \includegraphics[scale=0.6]{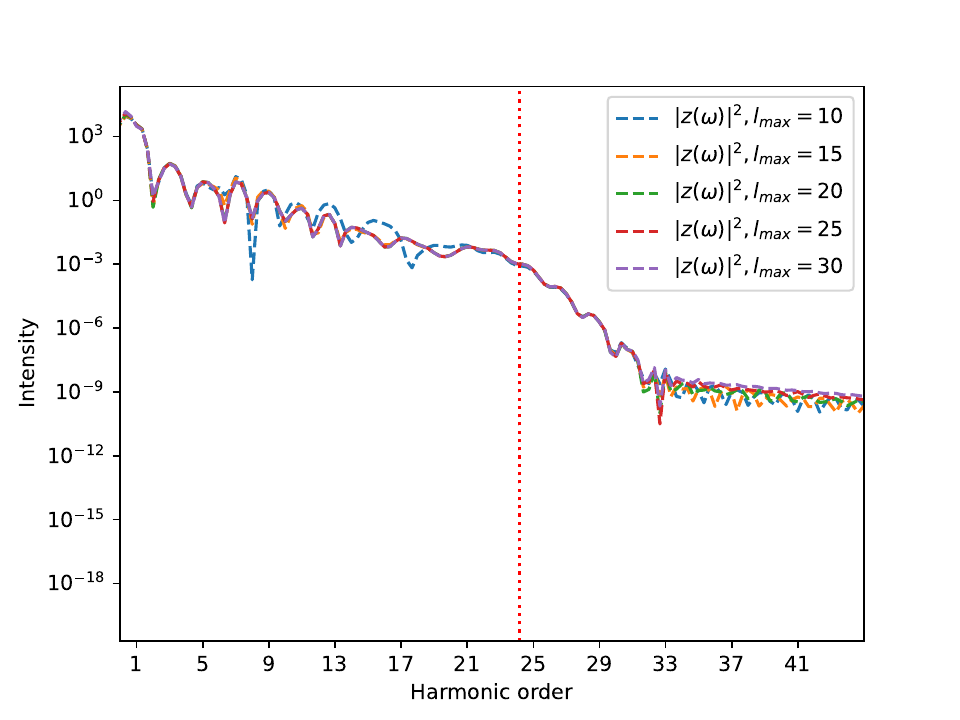}
\end{subfigure}
\begin{subfigure}[b]{0.6\textwidth}
   \includegraphics[scale=0.6]{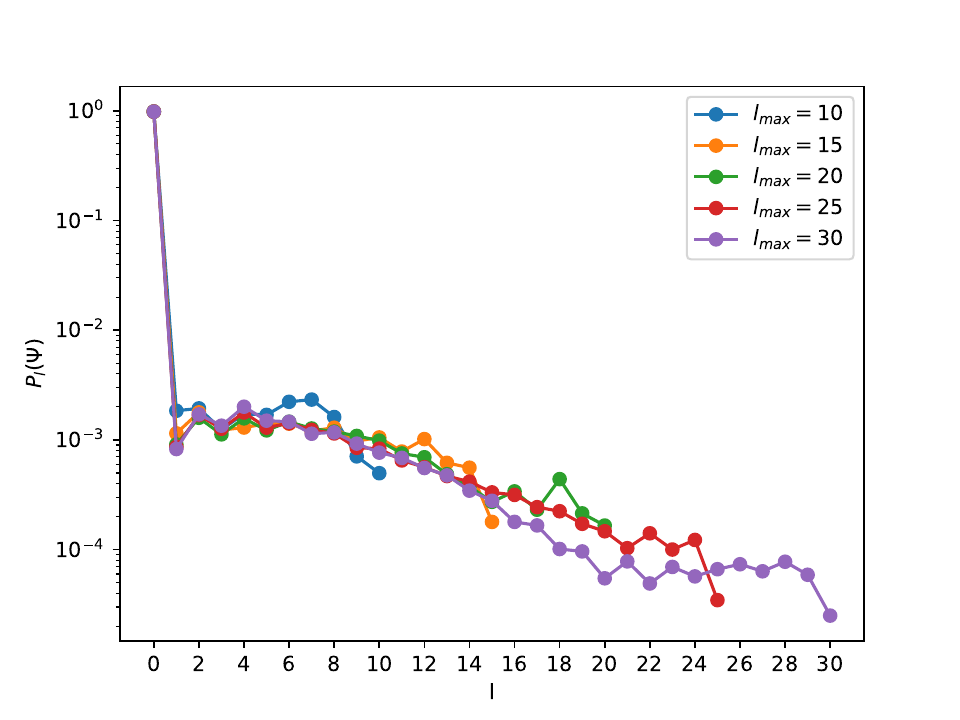}
\end{subfigure}
\begin{subfigure}[b]{0.6\textwidth}
   \includegraphics[scale=0.6]{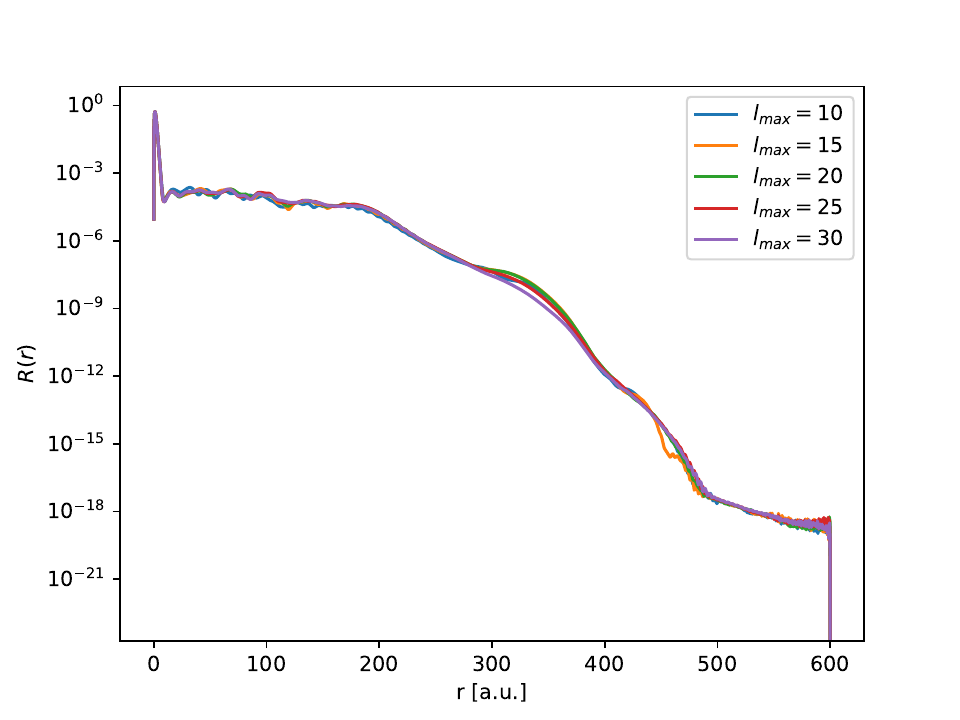}
\end{subfigure}
\caption[Description]{HHG spectrum, angular momentum and radial distribution at $t=T_f$ for $E_0 = 0.06\,\text{a.u.}$ using $r_{\text{max}}=600\,\text{Bohr}$ and $N=1200$.}
\label{fig:hhg_conv_006} 
\end{figure*}

\begin{figure*}
\centering
\begin{subfigure}[b]{0.6\textwidth}
   \includegraphics[scale=0.6]{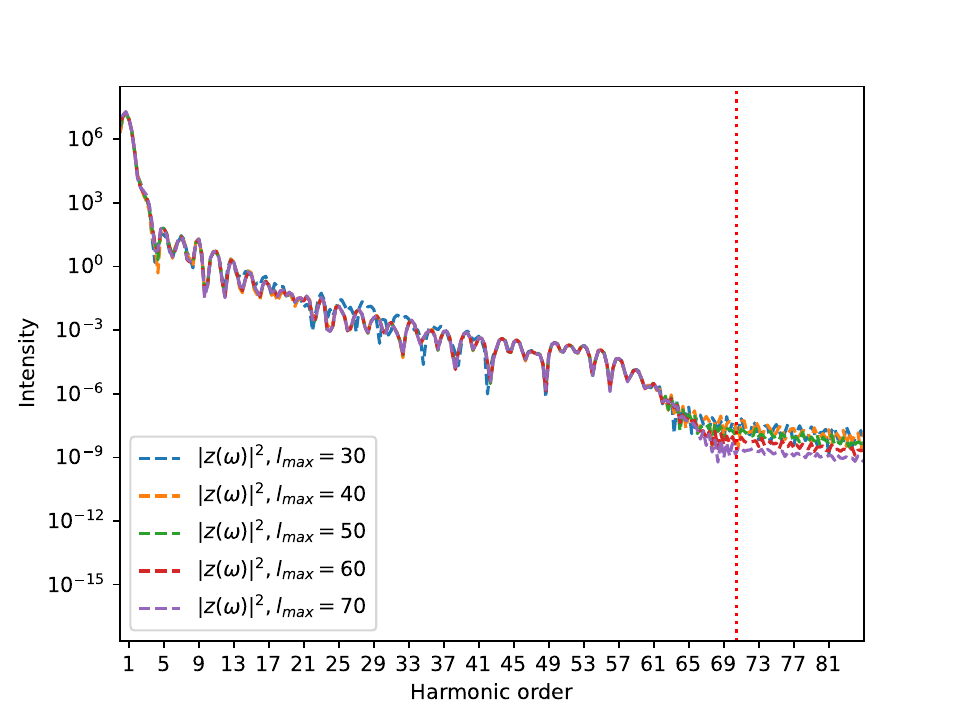}
\end{subfigure}
\begin{subfigure}[b]{0.6\textwidth}
   \includegraphics[scale=0.6]{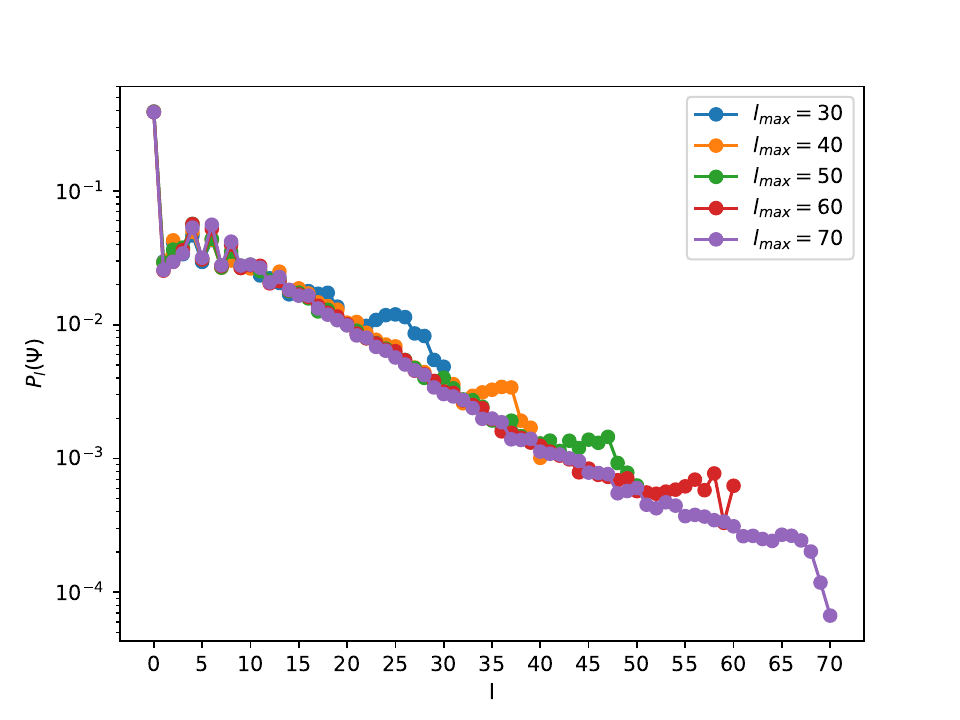}
\end{subfigure}
\begin{subfigure}[b]{0.6\textwidth}
   \includegraphics[scale=0.6]{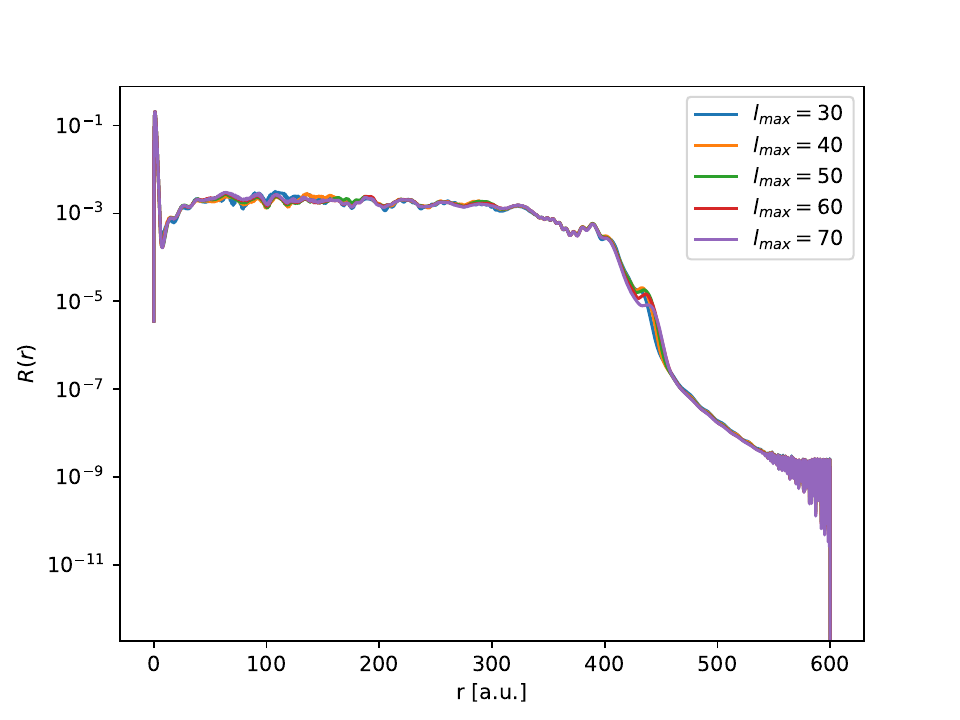}
\end{subfigure}
    \caption[Description]{HHG spectrum, angular momentum and radial distribution at $t=T_f$ for $E_0 = 0.12\,\text{a.u.}$ using $r_{\text{max}}=600\,\text{Bohr}$ and $N=1200$.}
\label{fig:hhg_conv_012} 
\end{figure*}


\end{document}